\documentclass[12pt]{iopart}
\usepackage{iopams}
\usepackage{graphicx,epsfig}
\usepackage{dcolumn}
\usepackage{bm}
\usepackage{color}

\newcommand{\be}{\begin{equation}}
\newcommand{\ee}{\end{equation}}
\newcommand{\bq}{\begin{eqnarray}}
\newcommand{\eq}{\end{eqnarray}}
\newcommand{\bc}{\begin{center}}
\newcommand{\ec}{\end{center}}

\begin{document}

\begin{flushright}
DAMTP-2010-29\\
DESY 10-049\\
MPP-2010-43\\
ICCUB-10-026
\end{flushright}

\title[Constraints on cosmic opacity and BSM physics from cosmological distances]{Constraints on cosmic opacity and beyond the standard model physics from cosmological distance measurements}

\author{Anastasios Avgoustidis$^{1}$, Clare Burrage$^{2}$, Javier Redondo$^{3}$, Licia Verde$^{4}$, Raul Jimenez$^{4}$}
{\it $^1$  Centre for Theoretical Cosmology, DAMTP, CMS, Wilberforce Road, Cambridge CB3 0WA, England--a.avgoustidis@damtp.cam.ac.uk\\
$^2$ Deutsches Elektronen Synchrotron DESY, Notkestrasse 85, D-22607 Hamburg, Germany--clare.burrage@desy.de\\
$^3$ Max Planck Institut f\"ur Physik, F\"ohringer Ring 6, D-80805, Munich, Germany\\
$^4$ ICREA {\rm \&} Institute for Sciences of the Cosmos (ICC), University of Barcelona, IEEC, Barcelona 08028, Spain--liciaverde,raul.jimenez@icc.ub.edu\\}

\date{\today}

\begin{abstract}
We update constraints on cosmic opacity by combining recent SN Type Ia data compilation with  the latest measurements of the Hubble expansion at redshifts between 0 and 2.  The new constraint on the 
parameter $\epsilon$ parametrising deviations from the luminosity-angular diameter distance relation ($d_L=d_A(1+z)^{2+\epsilon}$), is $\epsilon=-0.04_{-0.07}^{+0.08}$ (2-$\sigma$).  
For the redshift range between $0.2$ and $0.35$ this corresponds to an opacity $\Delta\tau<0.012$ ($95\%$ C.L.), a factor of 2 stronger than  the previous constraint. Various models of beyond the standard model physics that predict violation of photon number conservation contribute to the opacity and can be equally constrained. In this paper we put new limits on axion-like particles, including chameleons, and mini-charged particles. 
\end{abstract}



\section{\label{intro} Introduction}

Cosmological observations provide constraints on different distance measures: luminosity distance (as provided e.g., by supernovae), angular diameter distance (as provided e.g., by baryon acoustic oscillations) and even on the expansion rate or the Hubble parameter as a function of redshift $z$. 
Both luminosity distance and angular diameter distance are functions
of the Hubble parameter. While combining these measurements helps to
break parameter degeneracies and constrain cosmological parameters,
comparing them helps to constrain possible deviations from the assumptions
underlying the standard cosmological model (e.g. isotropy), or to directly constrain 
physics beyond the standard model of particle physics (e.g. couplings of photons 
to scalar or pseudo-scalar matter).

The Etherington relation \cite{Etherington1} implies that,  in a cosmology based on a metric theory of gravity,  distance measures are unique: the luminosity distance is $(1+z)^2$ times the  angular diameter distance. This is valid in any cosmological background where photons travel on null geodesics and  where, crucially, photon number is conserved.

There are several scenarios in which the Etherington relation would be violated: for instance we can have  deviations from a metric theory of gravity, photons not traveling along unique null geodesics,  variations of fundamental constants, etc.
In this paper we want to restrict our attention on violations of the Etherington relation arising from the violation of photon conservation. 

A change in the photon flux during propagation towards the Earth  will affect the Supernovae (SNe) luminosity distance measures but not the determinations of the angular diameter distance.  
Photon conservation can be violated by simple astrophysical effects or by exotic physics.
Amongst the former we find, for instance, attenuation due to interstellar dust, gas and/or plasmas. Most known sources of attenuation are expected to be clustered and can be typically constrained down to the 0.1\% level \cite{Menard, Bovy}. Unclustered sources of attenuation are however much more difficult to constrain.  For example, gray dust \cite{Aguirre}  has been invoked to explain the observed dimming of Type Ia Supernovae without  resorting to cosmic acceleration.

More exotic sources of photon conservation violation involve a
coupling of photons to particles  beyond the standard model of
particle physics.  Such couplings would mean that, while passing
through the intergalactic medium, a photon could disappear --or even (re)appear!-- interacting with such  exotic particles, modifying the apparent luminosity of sources.  
Here we consider the mixing of photons with scalars, known as axion-like
particles, and the possibility of mini-charged particles which have a
tiny, and unquantised electric charge.  
A recent review \cite{Jaeckel:2010ni} highlights the rich phenomenology of these weakly-interacting-sub-eV-particles (WISPs), whose effects have been searched for in a
number of laboratory experiments and astronomical observations. 
In particular, the implications of this particles on the SN luminosity have been described in a number of publications~\cite{Csaki2,Mortsellaxions,Mirizzi:2006zy,Burrage:2007ew,Ahlers:2009kh}. 

One of the most interesting features of these models is that the exotic opacity involved could  in principle ``mimic" the value of a non-zero cosmological constant inferred from SNe measurements. However, this possibility can already be excluded (at least in the simplest WISP models) by the absence of distortions in the CMB or the spectra of quasars for axion-like-particles, and by arguments of stellar evolution in the case of mini-charged particles. 

In this paper we use improved bounds on cosmic opacity to further constrain the existence 
of exotic particles which can couple to the photon. The rest of the paper is organised as follows.
In section 2 we update constraints on transparency from the latest available data.  
In section 3 we discuss the implications of this for axion-like particles and chameleons, and in 
section 4 we consider mini-charged particles. We then forecast, in section 5, how the constraints 
will improve with distance measures from future, planned and proposed, surveys. We conclude 
in section 6.  Sections 3 and 4 discuss in detail the motivation, modelling  and  regime of 
applicability of the beyond the standard model physics we consider. Readers with a more 
focused interest on cosmology may concentrate on the beginning of section 3, sub-sections 3.4, 
3.5 and figures 2, 3, 4, 5, 6.  Appendix A summarises the cosmologically-relevant results of 
sections 3 and 4.  

Precursors of this paper can be found in~\cite{bassettkunz1,bassettkunz2,Song:2005af,More,AVJ}.

\section{\label{update} An update on cosmic opacity constraints}

In reference \cite{AVJ}, the authors use Type Ia SN brightness 
data (namely the SCP Union 2008 compilation \cite{Union}) in combination
with measurements of cosmic expansion $H(z)$ from differential aging 
of luminous red galaxies (LRGs) \cite{JVST,SVJ05} to obtain constraints 
on non-trivial opacity, at cosmological scales.  The basic idea is to study 
possible violations from the ``Etherington relation'' \cite{Etherington1}, the 
distance duality between luminosity distance, $d_L$, and angular diameter 
distance, $d_A$:
\be\label{Etherington}
d_L(z)=(1+z)^2 d_A(z)\, .
\ee
This identity depends only on photon number conservation and local Lorentz 
invariance.  It holds for general metric theories of gravity, where photons travel 
along unique null geodesics.  Since Lorentz violation is strongly constrained for 
the low energies corresponding to optical observations \cite{Kostelecky:2008ts}, 
the study of possible violations of Eq.~(\ref{Etherington}) through SN 
observations directly constrains photon number violation.  Any such 
systematic violations can then be interpreted as an opacity effect in the 
observed luminosity distance, parametrised through a generic opacity 
parameter, $\tau(z)$, as:
\be\label{tau} 
d^2_{L,obs}=d^2_{L,true} e^{\tau(z)} \, .
\ee  

Note that our ``opacity'' can have in principle both signs. In 
other words, this parametrisation also allows for apparent \emph{brightening} 
of light sources, as would be the case, for example, if exotic particles were also emitted 
from the source and converted into photons along the line of sight \cite{Burrage:2007ew}. 
From Eq.~(\ref{tau}) it is clear that the inferred distance 
moduli for the observed SNe picks an extra term which is linear in $\tau(z)$:
\be\label{DMs}
 DM_{obs}(z)=DM_{true}(z)+2.5[\log e] \tau(z) \, .
\ee  

On the other hand, one can also use other determinations of distance 
measures, which are independent of $\tau$, to constrain possible deviations
from Eq.~(\ref{Etherington}).  This approach was initiated in reference 
\cite{More} (see also \cite{bassettkunz1,bassettkunz2,Uzan,LazNesPer} for related earlier 
work) where the authors used measurements \cite{PercivalBAO} of the baryon acoustic oscillation (BAO) scale at two redshifts, namely $z=0.20$ and $z=0.35$, to obtain a parameterization-independent upper-bound for the difference in  opacity between these two redshifts, 
$\Delta\tau<0.13$ at 95\% confidence.  In reference \cite{AVJ} this constraint was improved 
(and also extended over 
a wider redshift range, but for a  general parameterised form for $\tau$) by using, instead of measurements of the BAO scale at these two redshifts, measurements of cosmic expansion $H(z)$ 
from differential aging of LRGs at redshifts $z\lesssim 2$.  This method of distance determination 
relies on the detailed shapes of galaxy spectra but not on galaxy luminosities, so 
it is independent of $\tau$.  

In particular, the authors introduced a parameter $\epsilon$ to study deviations 
from the Etherington relation of the form:
\be\label{epsilon}
d_L(z) = d_A(z) (1+z)^{2+\epsilon} \, ,
\ee  
and constrained this parameter to be 
$\epsilon=-0.01^{+0.08}_{-0.09}$ at 95\% confidence.
Restricted to the redshift range $0.2<z<0.35$, where $\tau(z)=2\epsilon z + 
{\cal O}(\epsilon z^2)$, this corresponds to $\Delta\tau<0.02$ at 95\% confidence. 
In the following sections, we will apply similar constraints on different parametrisations
of $\tau$ which correspond to particular models of exotic matter-photon coupling, namely 
 axion-like particles (ALPs), chameleons, and mini-charged particles (MCPs).  

Before moving to these models, we briefly update the above constraint on $\epsilon$ 
using the latest $H(z)$ data \cite{SJVKS}, which include two extra data points at 
redshifts $z=0.48$ and $z=0.9$, as well as the latest determination of $H_0$ 
\cite{Riess09}.  Even though the addition of these two extra data points alone 
significantly improves the constraints of reference \cite{AVJ}, the effect of $H_0$ 
is also quite significant, because it acts as an overall scale in the distance measures, 
which is marginalised over a Gaussian prior, and the measurement error in this 
determination is about half of that of the HST Key Project determination 
\cite{HSTKey} used in \cite{AVJ}.  
 
\begin{figure}[h]
  \begin{center}
    \includegraphics[height=2.8in,width=3in]{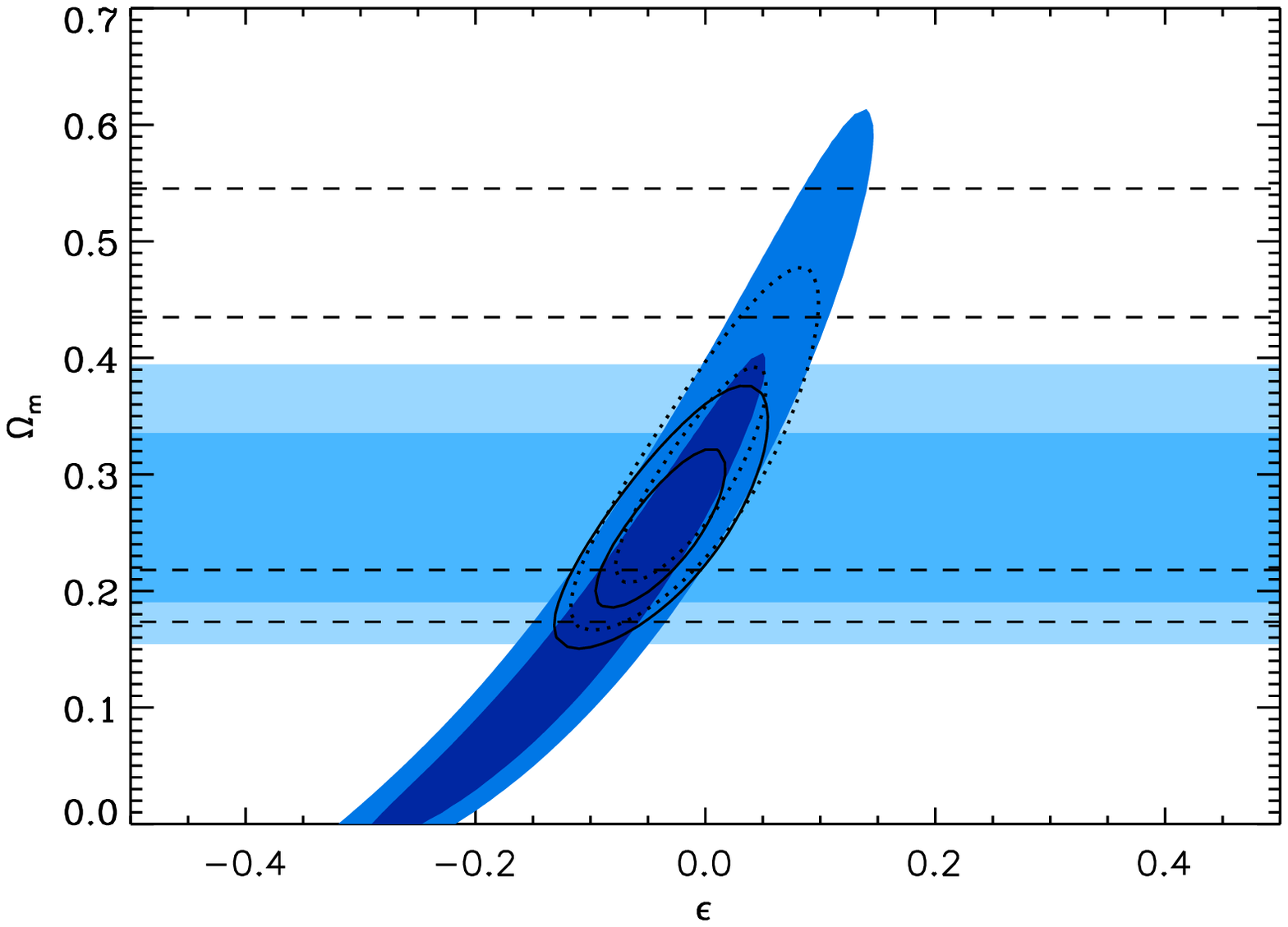}
    \includegraphics[height=2.8in,width=3in]{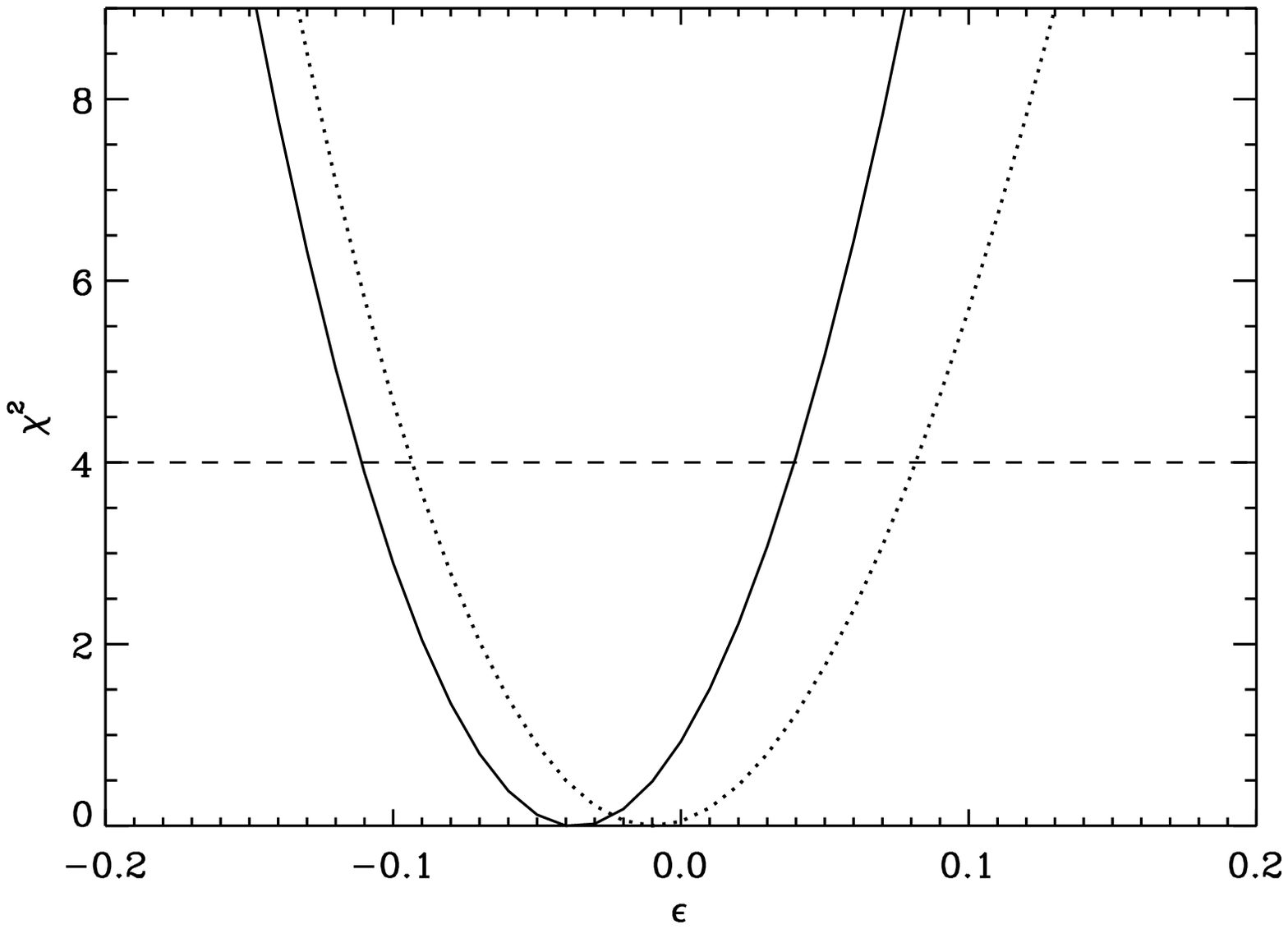} 
    \caption{\label{epsilon_constrs}  Updated constraints of reference 
                  \cite{AVJ}, using the latest $H(z)$ data \cite{SJVKS} and the 
                  Riess et al. determination of $H_0$ \cite{Riess09} in combination 
                  with the SCP Union 2008 SN Ia compilation.  
                  {\it Left:} Two-parameter constraints on the $\epsilon-\Omega_m$ 
                  plane.  Darker blue contours correspond to 68\% and 95\% 
                  confidence levels obtained from SN data alone, lighter blue 
                  contours are for $H(z)$ data, and solid line transparent contours 
                  are for joint SN+$H(z)$.  Previous $H(z)$ and joint SN+$H(z)$
                  from \cite{AVJ} are shown in dashed and dotted lines respectively.  
                  {\it Right:} One-parameter joint constraints on $\epsilon$ for the 
                  current analysis (solid line) and that of reference \cite{AVJ} (dotted
                  line). The dashed line shows the 95\% confidence level, $\Delta\chi^2=2$.}
  \end{center}
\end{figure}

Fig. \ref{epsilon_constrs} shows the updated constraints obtained using the 
above data in combination with the SCP Union 2008 Compilation \cite{Union} of type Ia 
Supernova data\footnote{Note that we decide to use the SCP Union 2008 supernova sample rather than the more recent  SDSSII sample \cite{kessler}, because the Union sample extends to higher redshift and is thus best suited to be combined with the $H(z)$ data.}, compared to the previous constraints of reference \cite{AVJ}.  On the left, the darker blue contours correspond to the (two-parameter) 68\% 
and 95\% joint confidence levels obtained from SN data alone, while lighter blue 
contours are the corresponding confidence levels for $H(z)$ data.  Solid-line 
transparent contours are for joint SN+$H(z)$ data.  For comparison 
we also show the previous $H(z)$ and  SN+$H(z)$ contours in dotted
and dashed lines respectively.  On the right we show one-parameter  (marginalized over 
all other parameters) constraints on $\epsilon$, again for the current analysis (solid line) 
and for that of reference \cite{AVJ} (dotted).  For the reader familiar with Bayesian 
methods, this plot corresponds to the posterior
\be\label{post_prob}
P(\epsilon|{\rm S},{\rm E})=\int_{\Omega_m}\int_{H_0} P(\Omega_m,H_0|{\rm E})
P(\epsilon,\Omega_m,H_0|{\rm S}) \, {\rm d}\Omega_m {\rm d}H_0 \,,
\ee
where $P(\Omega_m,H_0|{\rm E})$ and $P(\epsilon,\Omega_m,H_0|{\rm S})$ are
the posterior probabilities for the corresponding model parameters, given
the $H(z)$ ({\bf E}xpansion) and SN ($\bf S$upernovae) data respectively.
These are given by the likelihoods of the two data sets in the model parameters,
assuming Gaussian errors and using flat priors on all three parameters.  In 
particular, we have taken $\epsilon\in [-0.5,0.5]$, $\Omega_m\in [0,1]$ and 
$H_0\in [74.2-3\times 3.6, 74.2+3\times 3.6]$ (Riess et. al. \cite{Riess09}), 
all spaced uniformly over the relevant intervals, in a flat $\Lambda$CDM model.  
Similarly, the solid line transparent contours on the left plot of 
Fig. \ref{epsilon_constrs} correspond to taking only the integral over $H_0$ 
in the right hand side of Eq.~(\ref{post_prob}), yielding, therefore, the 
two-parameter posterior $P(\epsilon,\Omega_m |{\rm S},{\rm E})$.    

As seen in Fig. \ref{epsilon_constrs}, the improvement in these constraints is
significant.  The new result on $\epsilon$, marginalised over all other parameters, 
is $\epsilon=-0.04^{+0.08}_{-0.07}$ at 95\% confidence, which for redshifts between 
$0.2$ and $0.35$ (currently probed by BAO data), corresponds to a transparency (i.e., $\tau\ge 0$) 
bound $\Delta\tau<0.012$, a factor of two tighter than the result in reference \cite{AVJ}\footnote{Note that the data slightly favour negative $\epsilon$ (thus the much stronger constraint on a positive $\Delta\tau$), but 
only at $\lesssim$ 1-$\sigma$ level.}.  We now move on to study 
more general parametrisations of cosmic opacity, tailored for particular models of exotic 
matter coupled to photons.

\section{\label{cham}Axion-like Particles and Chameleons}
New scalar or pseudo scalar particles from physics beyond the standard model, here denoted as 
$\phi$, may couple to photons through
\be\label{Lscalar}
\mathcal{L}_{scalar}=\frac{1}{4M}F_{\mu\nu}F^{\mu\nu}\phi
\ee 
and
\be\label{Lpseudo}
\mathcal{L}_{pseudo-scalar}=\frac{1}{8M}\epsilon_{\mu\nu\lambda\rho}F^{\mu\nu}F^{\lambda\rho}\phi
\ee
where $M$ is the energy scale of the coupling (another widely used notation is 
$g_{\phi\gamma}=1/M$), $F_{\mu\nu}$ the 
electromagnetic field strength and $\epsilon_{\mu\nu\lambda\rho}$ the Levi-Civita
symbol in four dimensions.  
Such fields are collectively known as Axion-Like Particles (ALPs), as a coupling of the form (\ref{Lpseudo}) arises for the 
axion introduced by Peccei and Quinn (PQ) to solve the strong CP problem \cite{Peccei:1977hh}. 
Interestingly, these fields also arise naturally in string theory (for a review see \cite{Svrcek:2006yi}). 

Axions, or axion-like-particles, can arise from field theoretic extensions of the standard model as Goldstone bosons when a global shift symmetry, present in the high energy sector, is spontaneously broken. 
In the PQ axion case, this symmetry is colour anomalous and the explicit breaking makes the axion pick up a small mass.  This mass is, up to a model-independent constant, proportional to the coupling (\ref{Lpseudo}).
For a generic ALP, however, the mass is in principle independent of the strength of its coupling, 
and in particular can be zero if the releted shift symmetry remains intact.  That is, for instance, 
the case of Arions~\cite{Anselm:1981aw}, the orthogonal combination of the PQ axion, 
if there are \emph{two} independent colour anomalous shift symmetries.  

Chameleon scalar fields are another very interesting type of ALPs~\cite{Brax:2009ey}. 
They were originally invoked in \cite{Khoury:2003aq,Khoury:2003rn} to explain the current accelerated expansion of the Universe with a quintessence field which can couple to matter without
giving rise to large fifth forces or unacceptable violations of the
weak equivalence principle.  
The chameleon achieves this because its mass depends on the local energy density. 
The environmental dependence of the mass of the chameleon means that it
avoids many of the constraints on the strength of the coupling, which normally
apply to standard scalar and pseudo-scalar fields as they are derived
from physics in dense environments. For a more detailed discussion see \cite{Burrage:2008ii}.  
The cosomology of the chameleon was explored in detail in~\cite{Brax:2004qh}, the possibility of the chameleon coupling to photons was first discussed in~\cite{Brax:2007ak} and such a coupling was 
shown to be generic in~\cite{Brax:2009ey}.

The Lagrangian terms given above mean that ALPs can affect the
propagation of photons; in particular, if photons traverse a
magnetic field there is a non-zero probability that they will
oscillate into ALPs \cite{Raffelt:1987im}.   Notice however that only
photons polarized perpendicular (parallel) to the magnetic field mix
with scalar (pseudo-scalar) particles. Therefore, the interactions between photons 
and ALPs in the presence of a magnetic field not only imply that photon number 
is not conserved, but can also alter the polarization of the light beam.
Both effects have been exploited in many searches for ALPs both in the laboratory and in
astronomical observations, see~\cite{Jaeckel:2010ni} for a recent review. 

\subsection{Modelling the effects of ALPs}\label{modelingALP} 
The presence of ALPs will have an impact on observations of SNe if they 
are observed through intergalactic magnetic fields.  In particular, it will lead 
to changes in the observed SN luminosities, in a redshift-dependent way.  
Many different mechanisms have been proposed which give rise to intergalactic
magnetic fields, however we do not yet have convincing evidence from
observations that they exist.  A magnetic field coherent over the
whole Hubble volume is limited, by observations of the CMB and Faraday
rotation, to $B\lesssim 10^{-9}\mbox{ G}$ \cite{Barrow:1997mj,Blasi:1999hu}. 
Fields with shorter coherence lengths are also constrained.  In particular, 
fields coherent on scales $\sim 50\mbox{ Mpc}$ must satisfy $B\lesssim 6\times 
10^{-9}\mbox{ G}$, while fields coherent on scales $\sim\mbox{Mpc}$ must satisfy
$B\lesssim 10^{-8}\mbox{ G}$ \cite{Blasi:1999hu}.  To explain the origin
of galactic magnetic fields it is expected that intergalactic magnetic
fields with coherence lengths $\sim\mbox{Mpc}$ are needed \cite{Kronberg:1993vk}.

In a constant, coherent magnetic field the probability of a suitably
polarized photon
converting into an ALP after traveling a distance $L$ is given by 
\cite{Raffelt:1987im}:
\begin{equation}
P_{\gamma\rightarrow \phi}(L)= \sin^2(2\theta)
\sin^2\left(\frac{\Delta(L)}{\cos 2\theta}\right) \, ,
\label{probcon}
\end{equation}
where 
\begin{eqnarray}
\Delta(L)&=&\frac{m_{\rm eff}^2 L}{4\omega}\\
\tan 2\theta &=&\frac{2B\omega}{Mm_{\rm eff}^2} \,,
\end{eqnarray}
where $\omega=2\pi\nu$ is the photon energy, $B$ is the strength of the magnetic field and 
$m_{\rm eff}^2=|m_{\phi}^2-\omega_P^2|$, with $m_{\phi}$ the mass of the 
ALP and $\omega_P^2=4\pi^2\alpha n_e/m_e$ the plasma frequency of the
medium which acts as an effective mass for the photons ($\alpha$ is the fine structure constant, $n_e$
the local number density of electrons, and $m_e$ the mass of the
electron).  In what follows we restrict our attention to very light
fields, $m_{\phi}^2<\omega_P^2$, where observations of the opacity of
the universe have the most power to constrain the strength of the
coupling for the ALP to photons.  

However, the intergalactic magnetic field is not coherent from the earth to the supernovae.
We model its fluctuations using the {\it cell magnetic field} model, whereby we assume that 
the magnetic field is split up into a 
large number of equally sized domains.  Within each domain the magnetic field is constant, 
its strength being the same in all domains, while its orientation is chosen randomly.
The cell magnetic field model is the simplest choice we can make to
approximate the structure of astrophysical magnetic fields, and is
commonly used both in the study of ALPs and of astrophysical magnetic fields.  
A more accurate choice for a model of
the magnetic field would be to assume a power spectrum of its
fluctuations.  However, at high frequencies, $\Delta \ll \pi/2$,  the cell and power 
spectrum models give the same results, and at lower frequencies the cell model
captures all the qualitative features of ALP-photon interactions, but
underestimates the probability of conversion \cite{Davis:2009vk,Schelpe:2010he}.  
Therefore, using the cell magnetic field model will give rise to conservative
constraints. 

Clearly, we also need to know the plasma frequency of the intergalactic
medium along the line of sight. 
This is  quite a complicated issue because no measurements of the electron density are available in the large voids of the interstellar medium. 
A large-scale average can be easily inferred from the total amount of electrons determined by  CMB estimation of the baryon to photon ratio giving
$\omega_P\simeq 1.8\times 10^{-14}\, \mbox{eV}$ today, see e.g. \cite{Peebles:1994xt}. 
Note, however, that average values up to a factor of 15 smaller were considered plausible 
in~\cite{Csaki1}.  Since there is no easy way out of this conundrum, the accepted approach is to assume that $\omega_P^2$ is homogeneous and equal to the average value. To check the dependence of the results on this assumption, we will finally allow a range of a couple of orders of magnitude around the average.

As we are interested in the transparency of the Universe out to
redshifts $z\sim \mathcal{O}(1)$ we must also take into account the
redshift evolution of the environment that causes mixing between
photons and scalars.  Assuming the magnetic fields are frozen into the
plasma, their strength scales as $B(z)=B_0(1+z)^2$~\cite{Kronberg:1993vk} while 
$\omega(z)=\omega_0(1+z)$ and $\omega_P^2(z)=\omega_{P0}^2(1+z)^3$
(since it is proportional to the electron density).  Here, the
subscript $0$ indicates values in the present epoch.
The physical length of a magnetic domain scales as $L(z)=L_0(1+z)^{-1}$ as long as it is smaller than the Hubble radius.
Then the two parameters that appear in the probability of conversion (\ref{probcon}), $\theta$ and $\Delta$, redshift as
\begin{eqnarray}
\label{redshiftdependence}
\Delta(z)&=& \Delta_0 (1+z)\; ,\\
\tan 2\theta(z) &=& \tan2\theta_0\; .
\end{eqnarray}

There are two limits in which the expression for the conversion probability in one domain simplifies notably, the coherent and incoherent regimes. 
In the coherent regime, the argument of the sinus in~Eq.(\ref{probcon}) is smaller than 1 and, taking $\sin x\sim x$, the probability takes a very simple expression:
\begin{eqnarray}
P_{\gamma\rightarrow\phi}(z)&\approx& \left(\frac{B_0L_0}{2M}\right)^2(1+z)^2 \quad\quad (\rm coherent)\,.
\end{eqnarray}
On the other hand, if the argument of the sinus  -- which is energy dependent -- is very large, then a large number of oscillations will happen in a finite energy bin, which would average out the sinus to $1/2$. 
In this case we find
\begin{eqnarray}
P_{\gamma\rightarrow\phi}(z)& \approx & \frac{1}{2}\left(\frac{2B_0 \omega_0}{M \omega_{P,0}^2}\right)^2 \quad\quad\quad (\rm  incoherent) \,.
\label{weakprob}
\end{eqnarray}
These approximations are only valid for small values of $P_{\gamma\to \phi}$. 
Finally, note that from now on we will drop the subscript $0$ for today's values of the various 
parameters and make the redshift dependence explicit.

In the above limits, the redshift dependence is very simple and a system of axion-like particles and
photons can be evolved analytically through a large number of randomly oriented
magnetic domains. 
Let us introduce the notation $P$ for the transition probability $P_{\gamma\to \phi}$ in one 
domain today.  Assuming that the magnetic fields and intergalactic medium do not evolve 
with redshift, the photon survival probability was first computed in \cite{Grossman:2002by}. 
This is then exactly valid in the incoherent regime of Eq.~(\ref{weakprob}).  The finite 
probability of conversion gives rise to an apparent change in luminosity 
distance.  In particular, if photons are converted to ALPs along the line of sight, then 
the inferred and true and luminosity distance squared (cf Eq.~(\ref{tau})) will 
differ by a factor ${\cal P}(z)$, which in this case reads: 
\begin{equation}
{\cal P}(z)= A + (1-A) \exp\left(-\frac{3}{2}\frac{y(z)}{L}P\right) \; ,
\label{Pnoredshift}
\end{equation}
where $y(z)$ is the comoving distance to the source.  Physically, $\mathcal{P}(z)$ is the average probability that a photon emitted by a supernovae at redshift z is observed by us after traversing the magnetic fields 
in the intergalactic medium.  The above formula is valid for small $P$; in the 
case where $P$ is of order unity, one should replace $3P/2\to -\ln(1-3P/2)$.
We have allowed for an initial flux of axions $I_{\phi}(z_I)$ and defined\footnote {Very recently it was pointed out that this formula is actually an averaged formula over different realisations of the configuration of magnetic field domains along a line of sight~\cite{Mirizzi:2009aj,Burrage:2009mj}.  As such, it is in principle not valid for a single source, whose light only travels through a concrete realisation of the magnetic field domain structure. The authors of~\cite{Mirizzi:2009aj} calculated an analytical estimate for the dispersion around the mean -- although the distribution is non-Gaussian and in general asymmetrical so a meaningful confidence interval has to be computed numerically -- for $A=2/3$. Using $x=Py(z)/L$, the result reads  
\be
\delta {\cal P} = \sqrt{R -{\cal P}^2}\quad {\rm with }\quad R= \frac{49+50 \exp (-3x/2)+6 \exp (-5x)}{105}\,,
\ee
which falls in the range $\sim(0,1/(3\sqrt{5}))$, i.e. the dispersion is below $15\%$. 
In the Hubble diagram, this corresponds to a maximum dispersion of $0.2$ magnitudes, which is the typical dispersion in the SN data we use.
This otherwise suggestive fact, implies that confronting the observed with the predicted dispersion can only potentially constrain a regime of large values of $x$ -- what we will later introduce as strong mixing regime -- and most likely with small significance. 
We conclude that it is more promising to confront the mean value, given by Eq.~(\ref{Pnoredshift}), with observations.  In fact, as we shall see below, when constraining ``opacity" as a function of redshift, for every  reasonable redshift interval we may consider, there will be several data points corresponding to different supernovae at different positions in the sky, yielding an effective average; therefore in the following we can obviate the implications of the uncertainty in the value of $\delta P$.}
\begin{equation}
{\cal P}(z) = \frac{I_{\gamma}(0)}{I_{\gamma}(z_I)} \quad ; \quad A= \frac{2}{3}\left(1+\frac{I_{\phi}(z_I)}{I_{\gamma}(z_I)}\right) \ . 
\label{Pnoredshift_and_A}
\end{equation}

In the coherent regime, we wish to allow for additional effects due to the evolution of the 
magnetic fields with redshift.  The fluxes of scalars and photons at the end of the $n$-th 
domain are related to the fluxes at the beginning of the domain by
\begin{equation}
\left(\begin{array}{c}
I_{\gamma}(z)\\
I_{\phi}(z)
\end{array}\right)_{(n+1)}=\left(\begin{array}{cc}
1-\frac{1}{2}P(1+z)^2 & P(1+z)^2\\
\frac{1}{2}P(1+z)^2 & 1-P(1+z)^2
\end{array}\right)\left(\begin{array}{c}
I_{\gamma}(z)\\
I_{\phi}(z)
\end{array}\right)_{(n)}
\label{domain}
\end{equation}
and we want to multiply a large number of these matrices together. 
Usefully, the matrix  is diagonalisable by a redshift independent
transformation, so that,  after passing through $N$ domains, the photon survival probability can be easily shown to be
\begin{equation}
{\cal P}(z)= A + (1-A) \prod_{j=1}^{N} \left(1-\frac{3}{2}P(1+z_j)^2\right) \,, 
\end{equation}
where $\{z_j\}$ is a collection of redshifts in the range $(0,z)$, equally spaced in comoving length.
If the number of domains is large, we can approximate the product by an integral to get 
\begin{equation}
{\cal P}(z)= A + (1-A) \exp\left(-\frac{P}{H_0 L }\frac{H(z)-H_0}{
    \Omega_m H_0}\right) . 
    \label{Predshift}
\end{equation}
This resembles the expression for the photon survival probability when
the evolution of the background is neglected, equation (\ref{Pnoredshift}), 
but has a stronger $z$-dependence at large redshifts.   
Both formulae give the same results in the small $z$ regime, since $y(z)\simeq z/H_0$ 
and $(H/H_0-1)/\Omega_m \simeq 3 z/2$.

The redshift evolution of Eqs.~(\ref{Pnoredshift}) and
(\ref{Predshift}) is absent when $A=1$.  When this is the case, the
initial flux of photons and scalars is already thermalised
$I_{\gamma}(z_I)=2I_{\phi}(z_I)$ so that, on average, the effect of
photons converting into scalars is compensated by that of scalars converting
into photons, and thus no net effect of the mixing is seen.

In summary, the luminosity distance to a supernova is modified by an 
overall, redshift-dependent factor:
\be\label{dL_resc}
d_L\rightarrow d_L/\sqrt{{\cal P}(z)}\, . 
\ee
As mentioned in section \ref{update}, the effect of ALP-photon mixing 
described above can be interpreted as ``opacity'', generally of both signs, 
so that
\be\label{opacity}
\tau(z) =-\ln \mathcal{P}(z) \, .
\ee

We  now proceed to constrain the possibility of mixing between
scalars and photons, through their effects on cosmic opacity.
Again, there are two regimes in which analytical insight can be reasonably expected.

\subsection{Weak mixing}
\label{weak}
We begin by considering the case where the sum of the $\gamma\to\phi$ conversion probabilities 
in all the domains is smaller than unity ($NP_{\gamma\to \phi}\ll 1$). 
In this limit the effects of ALP-photon mixing are always small so it is known as weak mixing regime. 
When  the redshift evolution of the background is neglected, the probability of photon survival can be 
found \cite{Burrage:2008ii,Davis:2009vk,Schelpe:2010he} to be
\be
\label{P_cham_weak}
{\cal P}(z)=1-(1-A)\frac{3}{2}\left(\frac{y(z)}{L}\right) P \,.
\ee 
In the low redshift regime, both the redshift dependent and independent equations give the same opacity
\begin{equation}
\label{lowz}
\tau(z) =(1-A)\frac{3}{2}\frac{P}{H_0 L}z  \ .
\end{equation}
Note that this is exactly of the  form  used in section 2, $\tau(z) = 2 \epsilon z$. 
We can already obtain a first estimate of our bounds by using our improved constraint $\epsilon < 0.04$ at $95\%$ confidence.  Using $H_0\simeq 74.2$ km/s Mpc$^{-1}$ we obtain
\begin{equation}
P \lesssim 4 \times 10^{-5}  \frac{L}{\rm Mpc} \left(\frac{1/3}{1-A}\right)\;. 
\label{boundepsilon} 
\end{equation}

Note that $A\simeq 1$ cannot be constrained. As mentioned before, this situation corresponds 
to the initial flux of axions and photons having almost thermalized 
abundances (see Eq.~(\ref{Pnoredshift_and_A})).  A thermalized, or nearly 
thermalized, axion/photon flux has very small redshift dependence (since the mixing 
tends to thermalization, but this is almost complete before leaving the SN source) and thus distance measures have no constraining power.

Two further comments are in order. 
Our bound  of Eq.~(\ref{boundepsilon}) corresponds to the argument of the exponential taking values 
around $-0.081 z/(1-A)$. This corresponds to the start of the exponential regime for the larger redshifts, 
so it is  consistent with the Taylor expansion adopted, unless $A$ is fine-tuned to 1.
Finally, note that the bound on $\epsilon$ has been obtained  considering redshifts 
up to $z\sim 1.5$, while the validity of Eq.~(\ref{lowz}) is only ensured for small redshifts.
For this reason this bound should be considered as an order-of-magnitude estimation;  we present  an accurate,  numerical study, in section \ref{numerics}.

\subsection{Strong Mixing}
\label{strong}
The other analytically analysable limit of Eqs.~(\ref{Pnoredshift})
and~(\ref{Predshift}) is when the sum of the conversion probabilities in all domains is  
very large, $N P \gg 1$.  If the mixing between axion-like particles and photons 
is strong, then, on average, one-third of any initial flux will be converted  
into axion-like particles and two thirds into photons.  This can cause 
substantial changes to the apparent supernovae luminosities. It was
initially speculated \cite{Csaki2} that this could account for the dimming
of SNe without the existence of dark energy, however this has
now been excluded by observations of the CMB \cite{MirRafSerp}.  If 
there is a large initial flux of axion-like particles from SNe, then the 
SN images can be {\it brightened} by strong mixing between 
ALPs and photons in the intergalactic medium \cite{Burrage:2007ew}.

In the strong mixing limit  both equations~(\ref{Pnoredshift})
and~(\ref{Predshift}) give the same result, ${\cal P}\sim A$, that is, a constant independent of redshift. It is clear that our methods cannot constrain such a possibility, which would be equivalent to a change of the normalization of the SN luminosity.  Imposing $N P> 1$ at the closest SN redshifts up to
$z \sim 0.01$, we find
\begin{equation}
P> 0.015 \frac{L}{{\rm Mpc}} \, .
\end{equation}
We cannot constrain this range of probabilities  as the mixing is  too
strong  and the redshift dependence of the opacity is washed out.

\subsection{Numerical results}
\label{numerics}
In this subsection we turn our estimates into real constraints by means of a full likelihood analysis.
Note  that our results only depend on the conversion probability per comoving length $P/L$, so we cannot constrain $P$ and $L$ independently. 

\begin{figure}[t]
  \begin{center}
    \includegraphics[height=2.8in,width=3in]{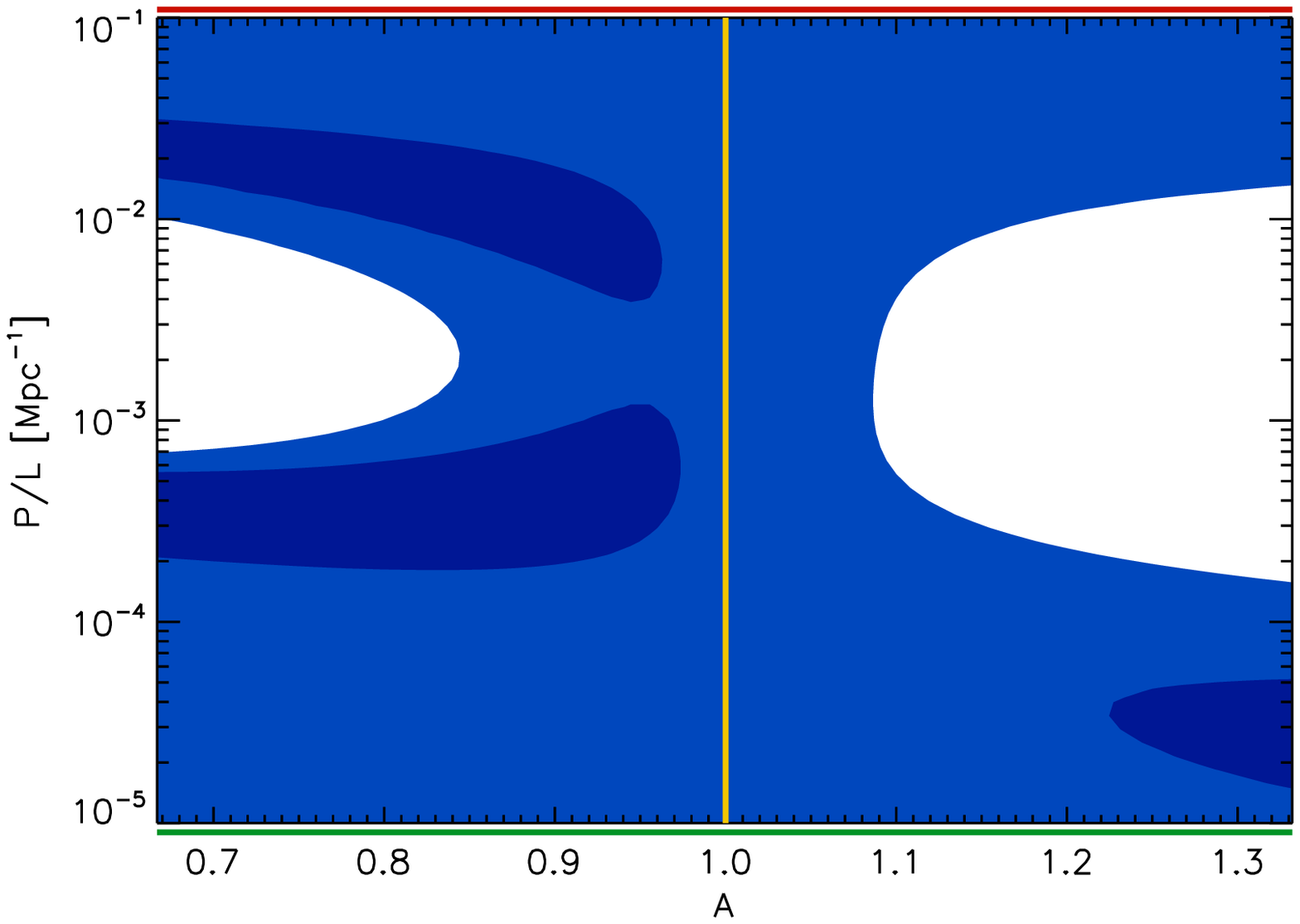}
    \includegraphics[height=2.8in,width=3in]{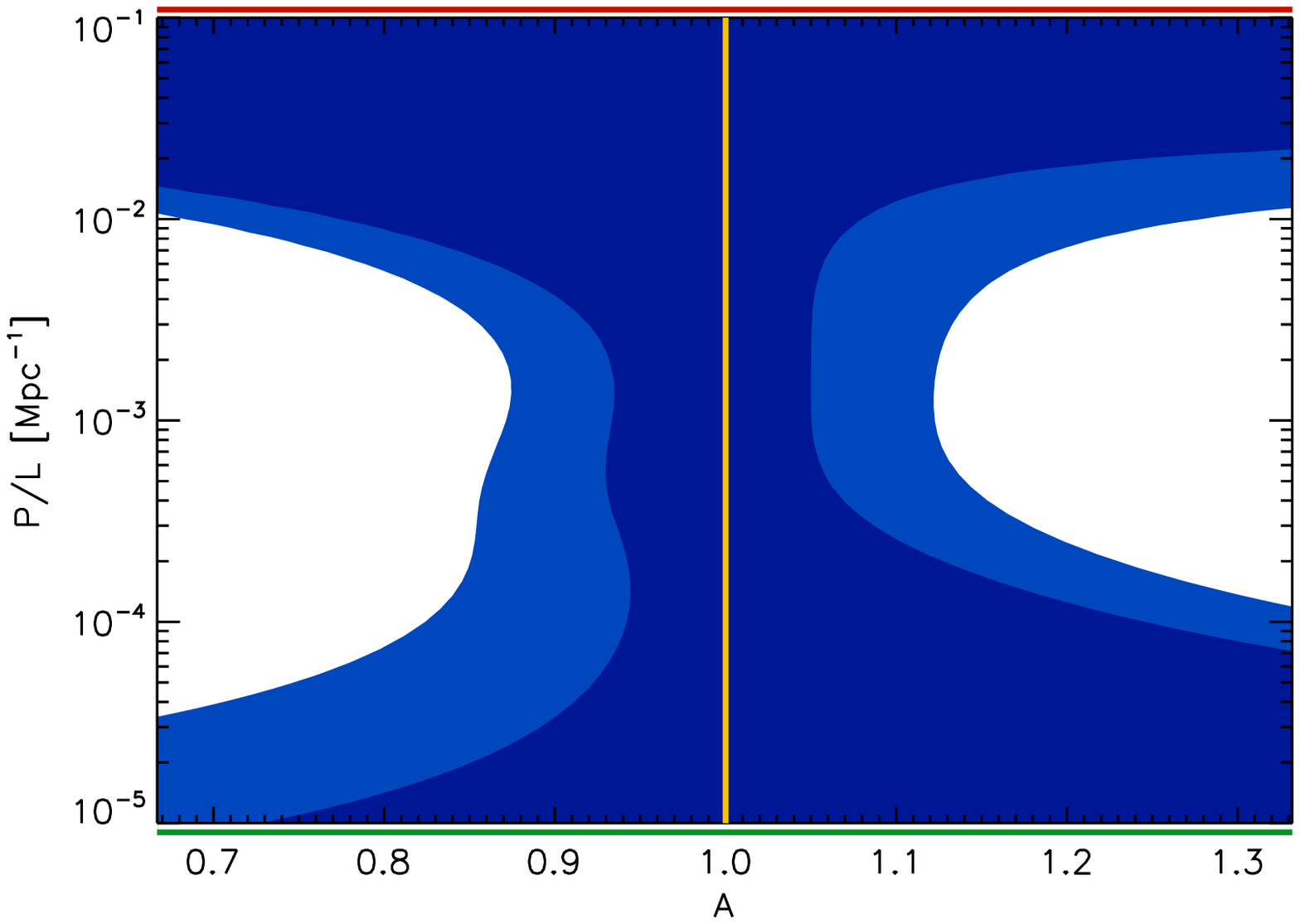} 
    \includegraphics[height=2.8in,width=3in]{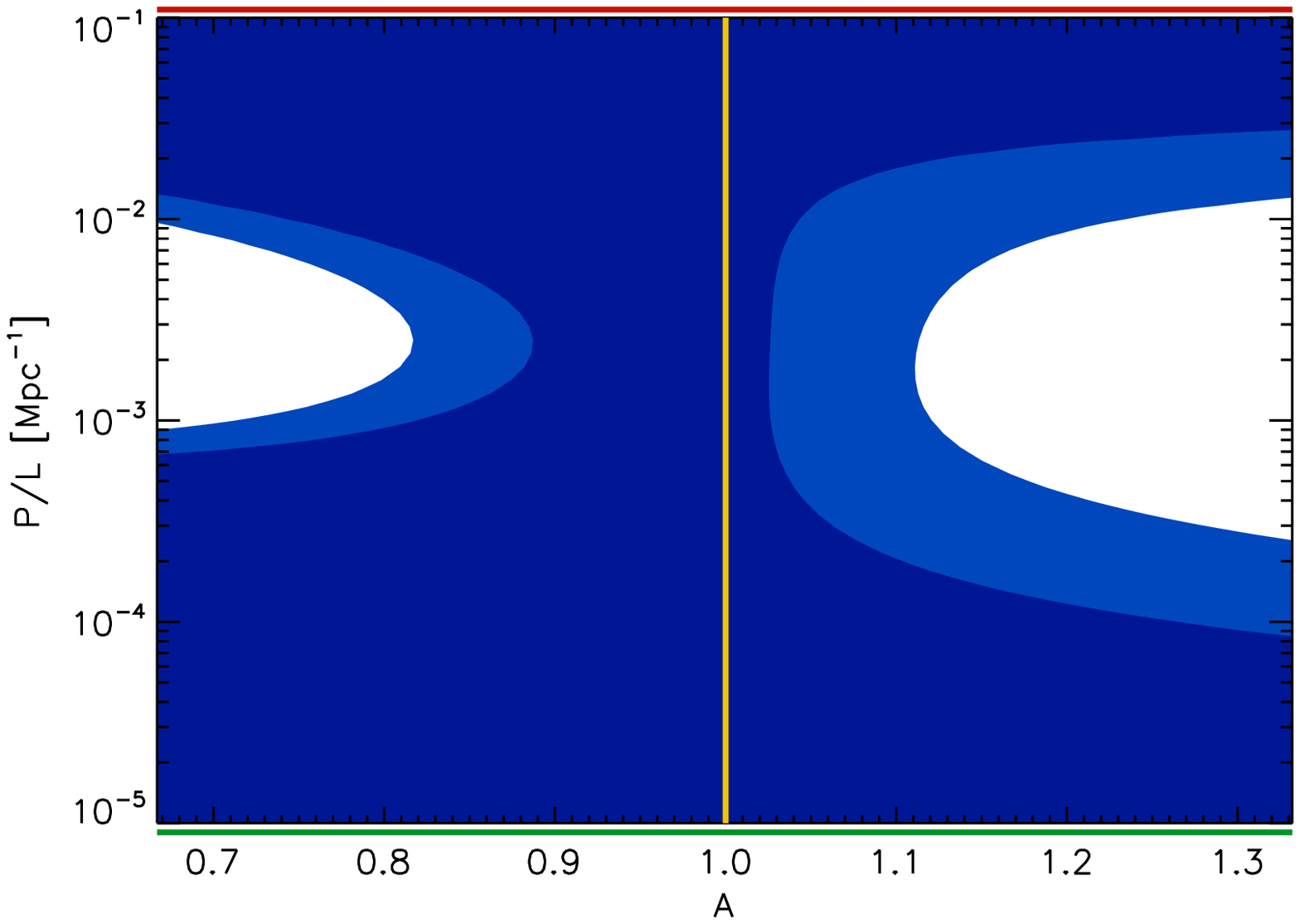}
    \includegraphics[height=2.8in,width=3in]{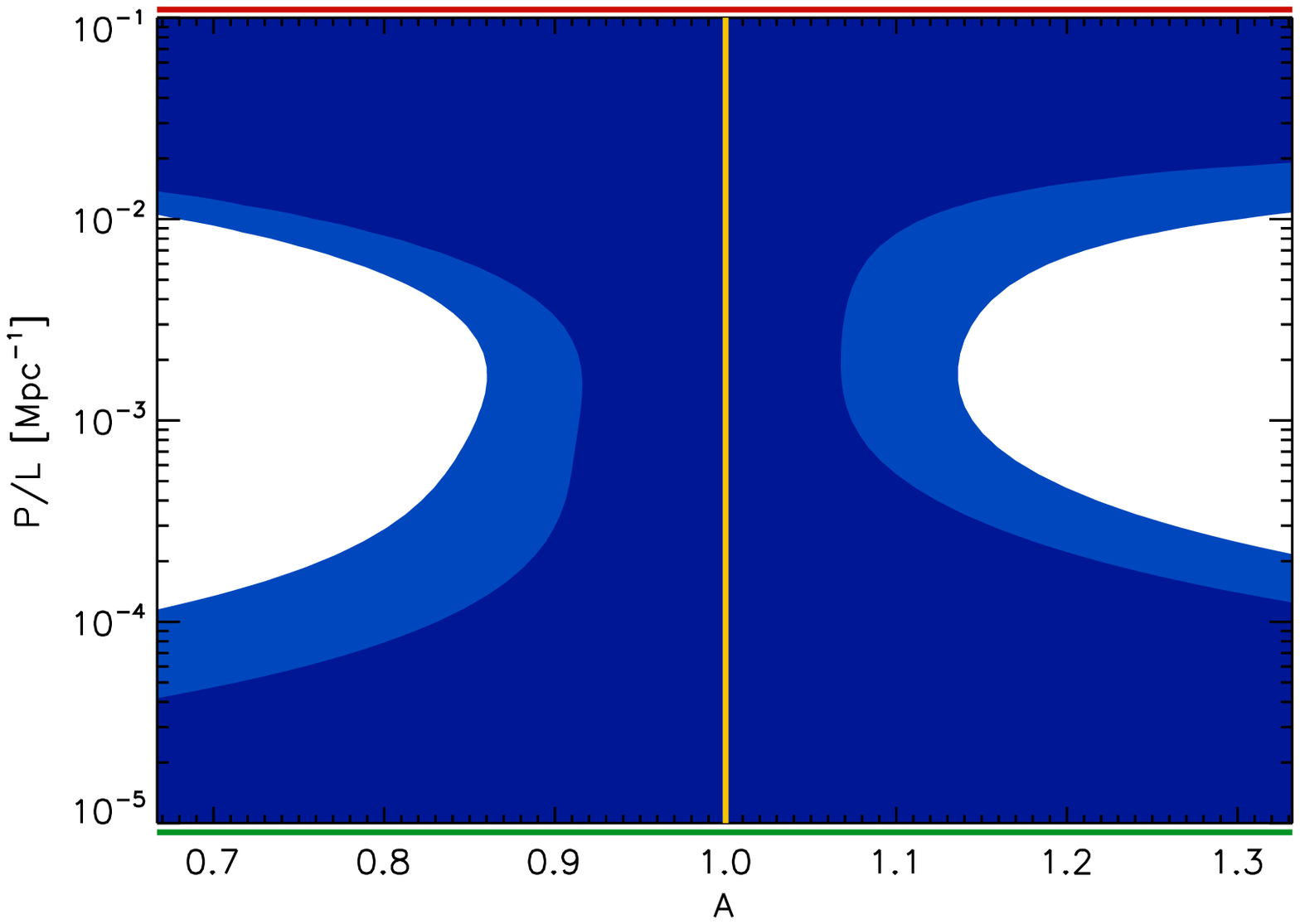} 
     \caption{\label{fig:redNored} Two parameter constraints in the $A-P/L$ plane for general ALPs considering redshift dependence of the background (upper panels) and neglecting it (lower panels). Contours represent the  $68\%$ (dark) and $95\%$ (light) confidence levels. For the left panels we have used SN data only, while in the right panels we show the joint SN+$H(z)$ analysis. The contours are marginalized over cosmologies and $H_0$.  While redshift dependence introduces ${\mathcal O}(1)$ effects, which change the structure of the $68\%$ CL contours, the $95\%$ CL contours are very similar in the two cases.}
  \end{center}
\end{figure}

In Fig.~\ref{fig:redNored} we show our results marginalized over $\Omega_m$ and $H_0$.  The dark and light contours represent $68\%$ and $95\%$ joint confidence levels respectively, using the SN data only (left) and joint SN$+H(z)$ data (right).  In
the upper panels, we have used Eq.~(\ref{Predshift}), thus taking into account the redshift dependence 
of the background, while in the lower panels we used Eq.~(\ref{Pnoredshift}), ignoring these effects.     

The strong and weak mixing limits described in sections \ref{strong} and
\ref{weak} are clearly visible in Fig. \ref{fig:redNored}.  We represent them 
schematically by the green and red lines below and above the boundaries 
of our plots respectively. 
For very small conversion probability $P\lesssim 10^{-5}$ we are in
the weak mixing limit and our constraints become weak because of the
lack of photon-ALP mixing. 
On the other extreme, when the probability is very strong ($P\gtrsim
{\rm few}\times 10^{-2}$) the photons and axions mix until thermalization 
and the redshift dependence of the opacity is lost, so, again, our constraints 
become weak.  Finally we can also observe that a band around $A=1$ (yellow line) 
again cannot be constrained. $A=1$ means an equilibrated photon-ALP flux from the SNe such that photon$\to$ALP conversions are compensated with the reverse process, making the photon number constant, i.e. redshift-independent.

The first notable feature is that, as expected from 
our estimations in the first section, our 2-$\sigma$ bounds (white regions) are significantly 
improved when including the $H(z)$ data (right panels) compared to using SN data alone (left panels). 
The improvement is particularly visible in the weak mixing regime.
This will show more clearly in the constraints 
on the $P/L-\Omega_m$ plane, which we will present below.  We shall comment on this 
improvement separately in each case of interest.

The second effect is the importance of including the redshift dependence of the magnetic fields, 
the photon frequency and the physical length of domains, that is, the difference between using 
Eqs.~(\ref{Pnoredshift}) and~(\ref{Predshift}).  The $95\%$ C.L. constraints are improved only 
slightly by including the redshift dependence (upper panels), compared to a
background that does not evolve (lower panels), the exception being
the factor of $3$ improvement at $A\simeq 2/3$ in the SN+$H(z)$ analysis (right panels).

A peculiar feature appears in the upper left plot, corresponding to a redshift-dependent background and using SN data only. There, the $68\%$ C.L. contour shows not one, but three  regions where photon-ALP mixing improves the fitting of the data compared to  a standard $\Lambda$CDM cosmology.  
The lower-left region corresponds to the parameters invoked by Csaki {\em et al.}~\cite{Csaki2} to explain SN dimming without cosmic acceleration. 
It is remarkable that the data show this preference much more sharply in the redshift-dependent background case given by Eq.~\ref{Predshift} than in  the simpler redshift-independent case used in~\cite{Csaki2}  (Eq.~\ref{Pnoredshift}). 
As we show in the next section, our numerical analysis implies that this island in parameter space corresponds to $\Omega_m<1$,  which, in the absence of a cosmological constant, implies a non-flat geometry.  Most importantly, the joint SN+$H(z)$ analysis (upper right panel) rules out this region.

The preferred region at large $A$ and small $P/L$ also deserves some comments.
A value of $A$ greater than unity produces an increase of the SN luminosity with redshift because the SNe would shine more ALPs than photons. In Fig.~\ref{epsilon_constrs} we have shown that, because of the pronounced degeneracy, SN data alone show a slight preference for this scenario since the $68\%$ C.L. contour is almost completely located at negative values of our opacity parameter $\epsilon$.
In the joint  $SN$+$H(z)$ analysis, the trend is softened but, still, small negative values (slight SN brightening) are slightly preferred, even though this is  not statistically significant. Therefore, in the remaining panels of Fig.~\ref{fig:redNored} this possibility cannot be excluded, but  the statistical preference for this region  decreases compared to the rest of the allowed parameter space.
The fact that photon-ALP mixing alleviates the tension between SN data and standard rulers was already pointed out in~\cite{Burrage:2007ew}. 
The value of $P/L\sim 2\times 10^{-5}$ Mpc$^{-1}$ implies that the required effect on the opacity is small. In the weak mixing limit, the opacity is proportional to $(1-A)P/L$ (see Eq.~(\ref{lowz})), making $A$ and $P/L$ correlated. This region is therefore expected to extend further right and down.  Notably, these small values of $P/L$ are not excluded by other arguments (see next section), although it seems difficult to conceive a model where $A$ is sufficiently large.

\subsection{Constraints}
We can now interpret our results in terms of constraints on the physical parameters of our Lagrangian. 
There are several cases to consider.

\subsubsection{Axion-like-particles:} 

\begin{figure}[t]
  \begin{center}
    \includegraphics[height=2.8in,width=3in]{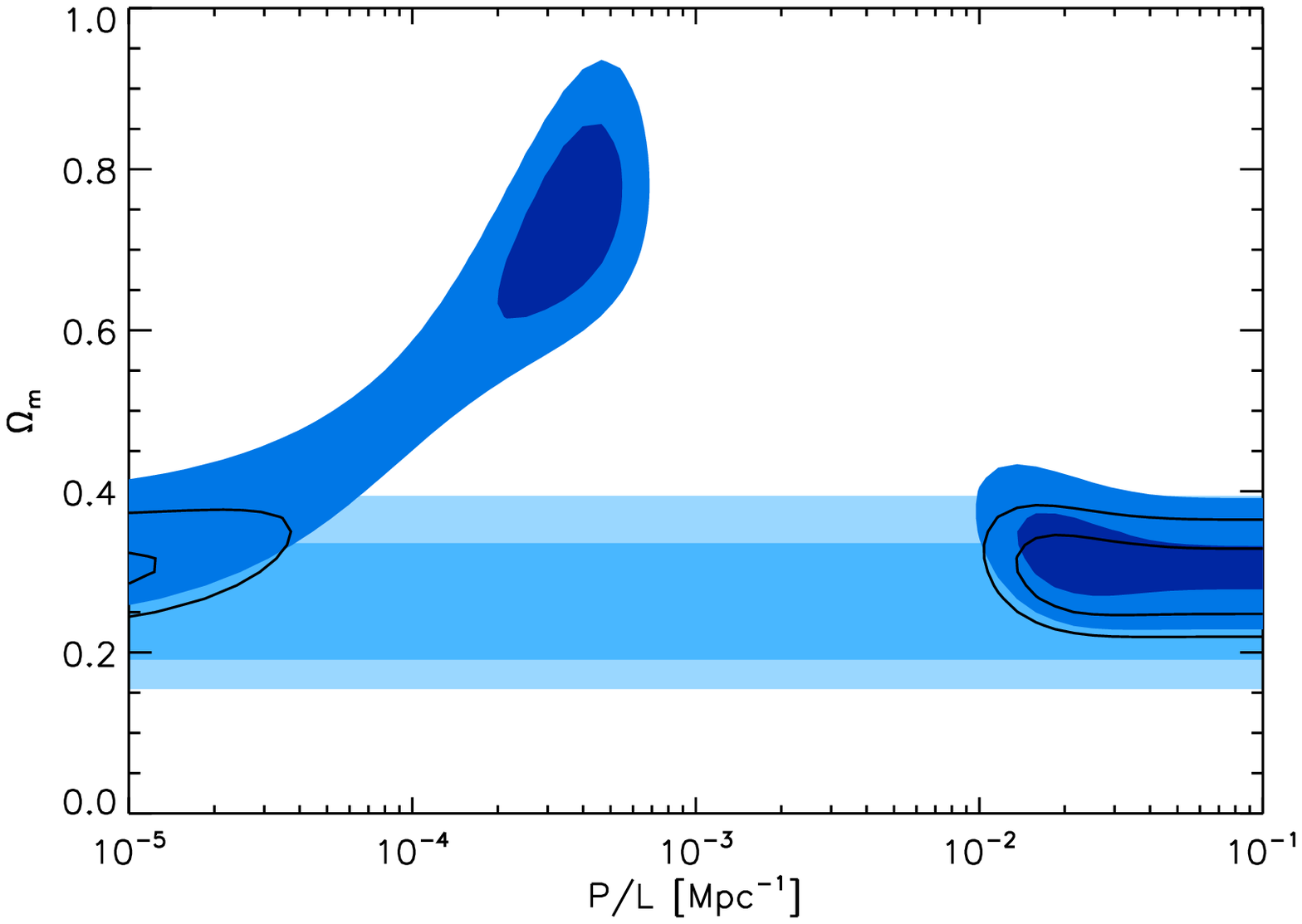}
    \includegraphics[height=2.8in,width=3in]{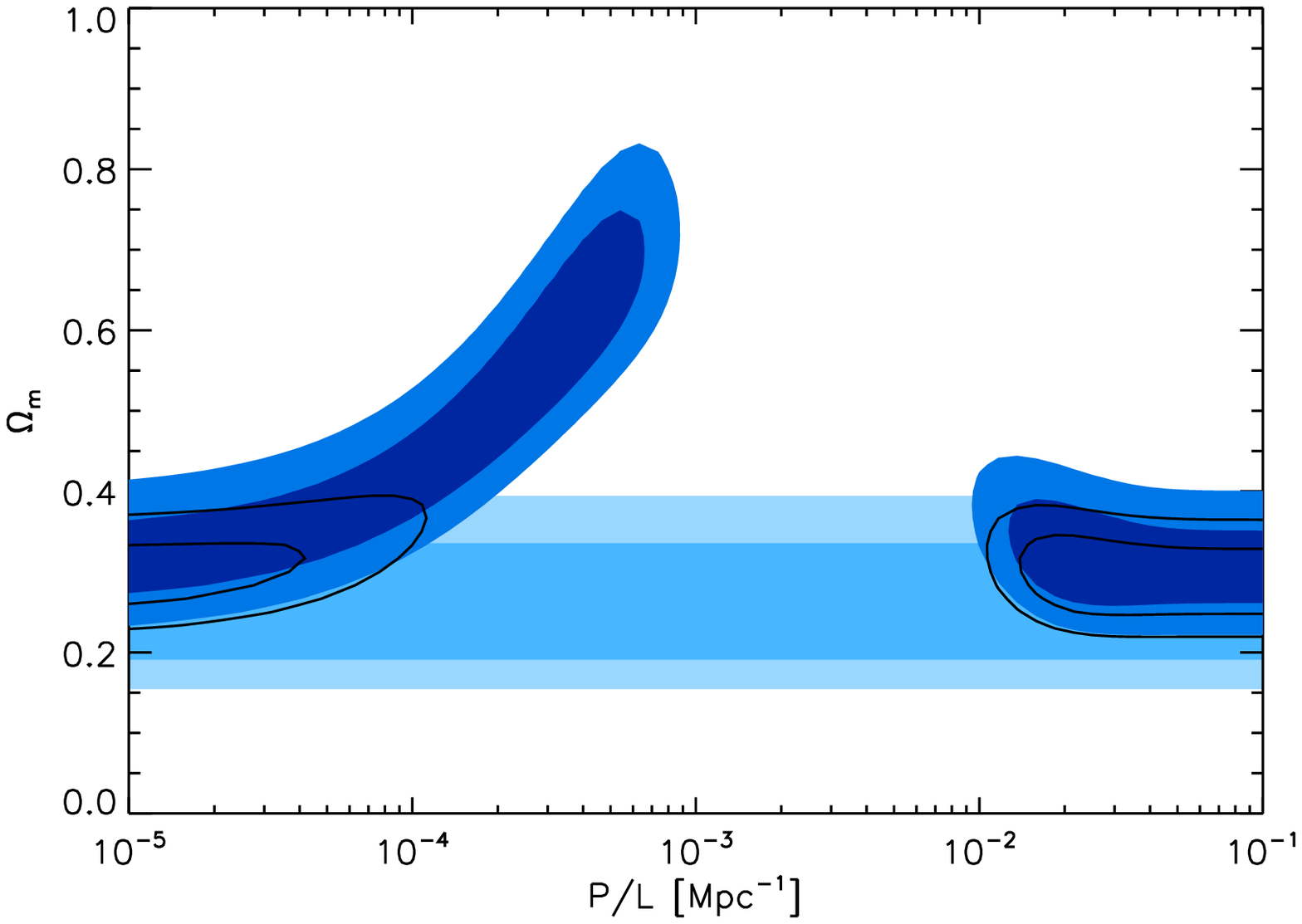}
    \caption{ \label{POmm} Confidence levels (68\% and 95\%) on the 
                  $P/L-\Omega_m$ plane for the simplest axion-like-particle model ($A=2/3$).
                  The small and large $P$ regions correspond to the weak and strong mixing regimes 
                  respectively.  Dark blue contours show constraints from SN data only, light blue 
                  from $H(z)$ data, and solid line contours from joint SN$+H(z)$. In the left panel, 
                  the redshift dependence of the background is taken into account, while in the right 
                  panel these affects are ignored.}
  \end{center}
\end{figure}

If ALPs have no other interactions than the two-photon coupling, then they were shown 
to contribute very little to the SN luminosity \cite{Grossman:2002by}, corresponding to the 
case $A\simeq 2/3$.  It is evident from Fig.~\ref{fig:redNored} that only a range of
conversion probabilities around $P/L \sim {\cal O}(10^{-3})$ can be excluded.
Fig.~\ref{POmm} shows 1- and 2-$\sigma$, two-parameter likelihood contours 
on the $P/L-\Omega_m$ plane after marginalization over $H_0$.  Note that even the 
SN constraints alone (dark blue contours) rule out this model as an alternative to a 
cosmological constant, at greater than $3$-$\sigma$ significance.  However, there is 
still significant degeneracy in the weak mixing regime, and a value for $\Omega_m$ greater 
than $0.8$ is still allowed at the $2$-$\sigma$ level, this constraint being slightly weaker when 
the background redshift dependence is taken into account (left).  
Including the $H(z)$ data (light blue contours) breaks this degeneracy, yielding strong joint constraints in $P/L-\Omega_m$
(solid line contours). 

We can now translate the bounds on $P/L$ into bounds on the strength
of the  ALP coupling to photons. 
Since the coupling always appears multiplied by the magnetic field (which is also unknown) we find convenient to quote bounds on the combination $B/M$. Let us also define appropriately normalised 
values of  the magnetic field strength $B$ and the energy-scale of the the  axion-photon coupling $M$, 
as 
\begin{equation}
B_{\rm nG} = \frac{B}{1\rm nG } \quad \quad ;\quad \quad  M_{\rm 10} = \frac{M}{10^{10} \rm GeV }\,.
\end{equation}

\begin{figure}[b]
  \begin{center}
    \includegraphics[width=3in]{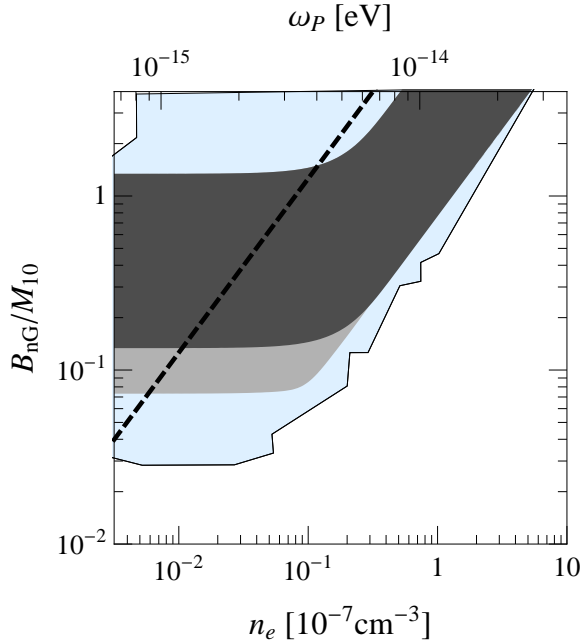}
     \caption{\label{fig:boundALP} 
     Bounds for the product of ALP coupling times magnetic field $B_{\rm nG}/M_{10}$ as a function of the average electron density or corresponding plasma frequency. 
     The gray regions are excluded at $95\%$ C.L. for $L=1$ Mpc in the redshift-independent (dark) and dependent (light) cases.
     Also shown are the regions constrained from CMB~\cite{MirRafSerp} (above the dashed line), which dominate at low $n_e$, and QSO~\cite{Ostman:2004eh} (blue) spectral distortions.}       
  \end{center}
\end{figure}

In Fig.~\ref{fig:boundALP} (left) we show our constraints for the case $L=1$ Mpc as a function of the uncertain value of the average electron density, or, equivalently, the plasma frequency.  
To get rid of the oscillations of Eq.~(\ref{probcon}), which
not only will be averaged out by energy binning but also by small fluctuations in the sizes of the domains and the values of the plasma frequencies, we propose the substitution $\sin^2 x\to (1-\exp(-2x^2))/2$, which reproduces the coherent and incoherent limits. The exclusion limit is a horizontal band which bends upwards around $n_e\simeq 0.2 \times 10^{-7}$ cm$^{-3}$.
The horizontal part corresponds to the coherent case, where the $n_e$ dependence drops out of $P$, while the diagonal  band corresponds to the incoherent case where $n_e$ suppresses $P$.
Note that the average electron density today is $n_e\simeq 10^{-7}$ cm$^{-3}$, near the transition 
between the two regimes.

In the same figure, we have also reproduced the constraints of Mirizzi {\em et al}~\cite{MirRafSerp} from distortions of the CMB  (region above the dashed line) and those of \cite{Ostman:2004eh} (blue region) from QSO spectra (see also footnote~2 in~\cite{MirRafSerp} and \cite{Mortsell1, Mortsell2}).
For $n_e\gtrsim 10^{-9} {\rm cm}^{-3}$ our bounds are stronger than the CMB ones while still competitive with the QSO bounds.  Our approach provides a complementary, independent way to  obtain these constraints. Each of the three approaches reported in the figure (especially the present constraint and the QSO one) is affected by different, unrelated, systematics: their agreement promotes one's confidence in 
these results.

\subsubsection{Chameleons:}
 Unlike the simple ALP case studied above, the chameleon Lagrangian
 contains non-linear self-interactions of the scalar field in order
 that the mass of the scalar may become dependent on the density of
 its environment.  This introduces the possibility of  having $A\neq 2/3$. 

There are three possible cases: i) either $A\simeq 2/3$ because few
chameleons are produced in the SN, ii) $A-1\ll 1$ because they
interact so strongly in the SN that photons and chameleons thermalize
their fluxes within the SN, and iii) the intermediate case where a
significant number of scalar particles are produced in the interior of
the SN, but, yet, photon-chamelon interactions are not strong enough 
to thermalise the chameleon population with that of the photons before 
they leave the SN.

\begin{figure}[t]
  \begin{center}
    \includegraphics[height=2.8in,width=3in]{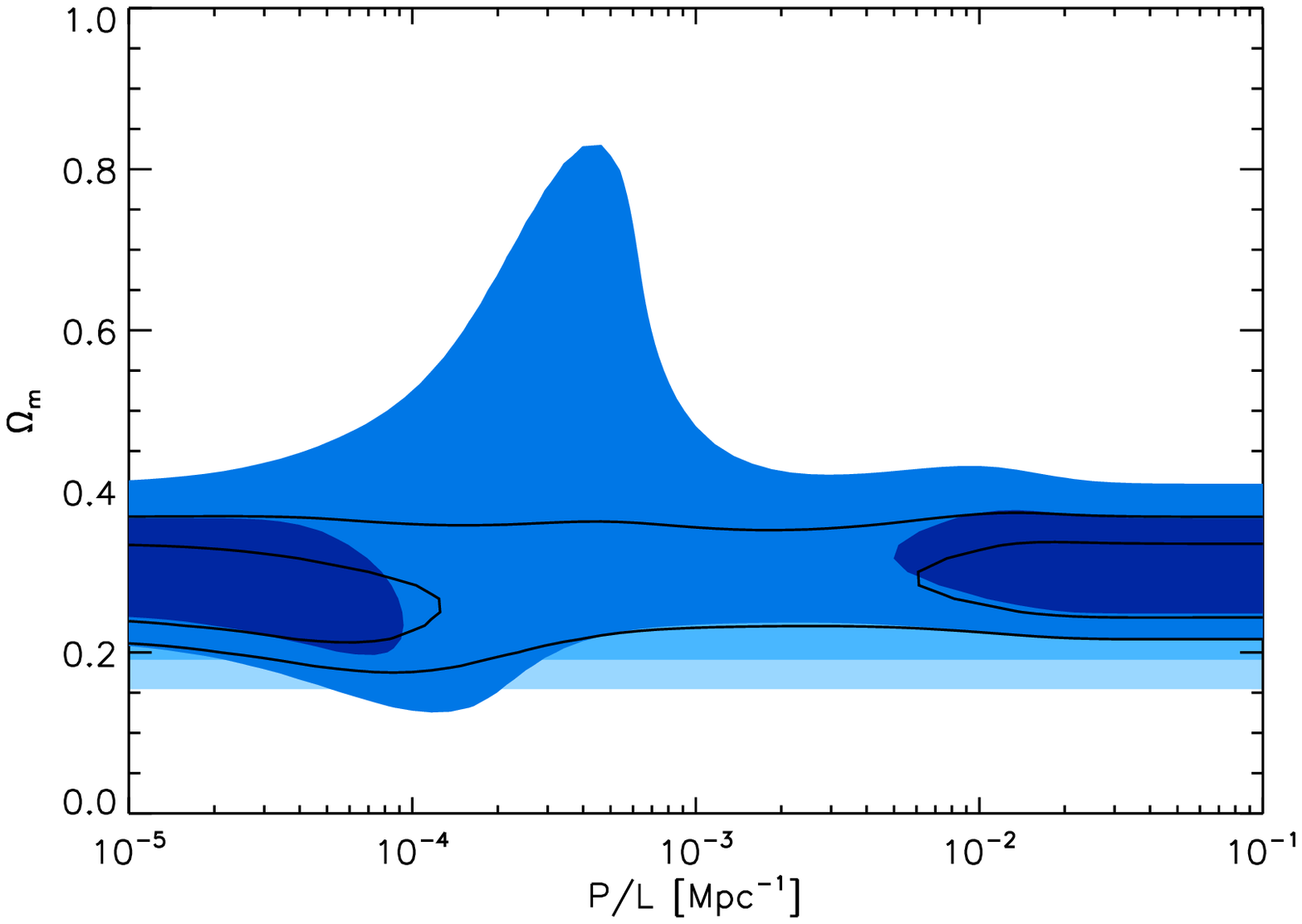}
    \includegraphics[height=2.8in,width=3in]{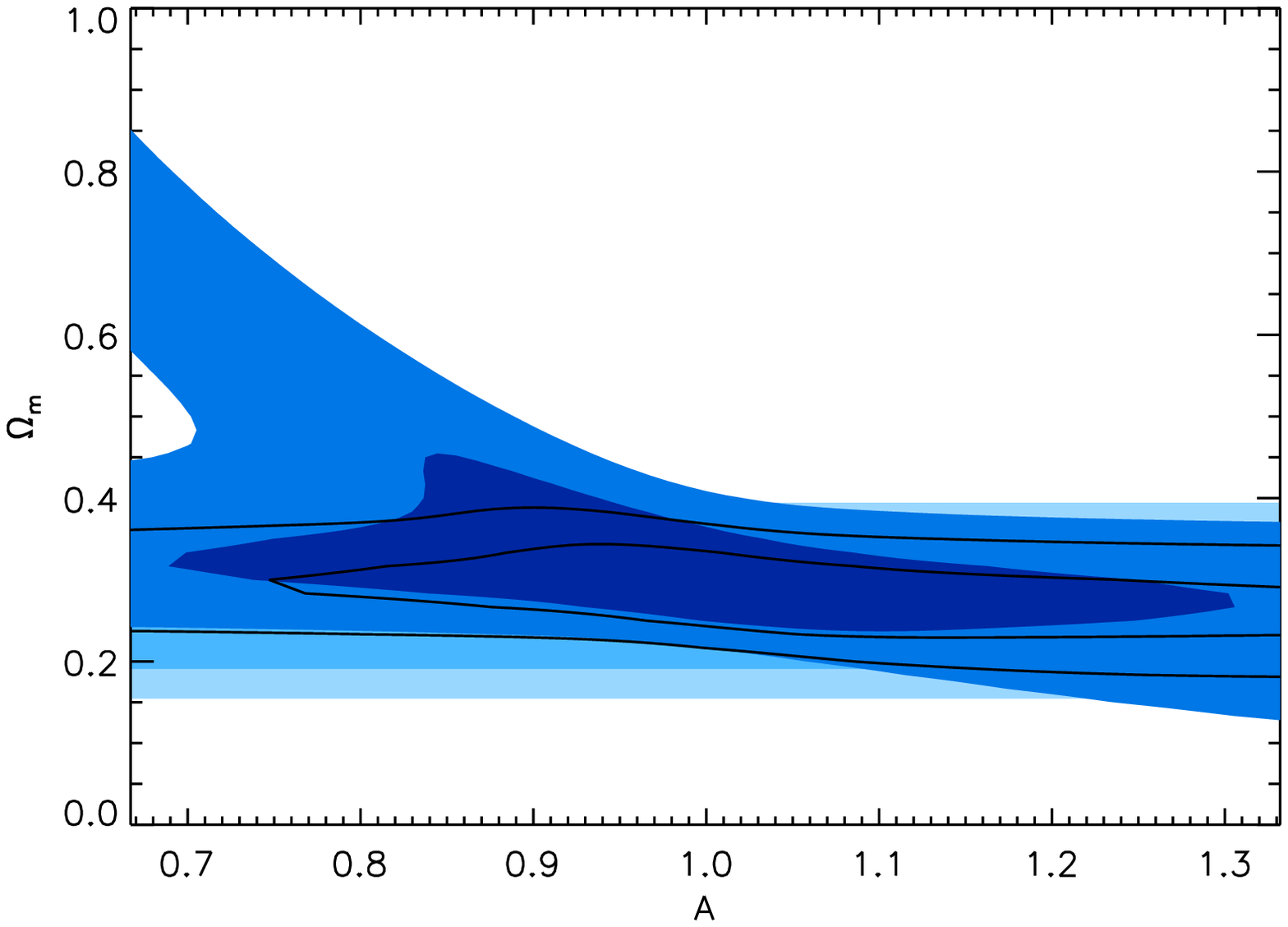}
    \includegraphics[height=2.8in,width=3in]{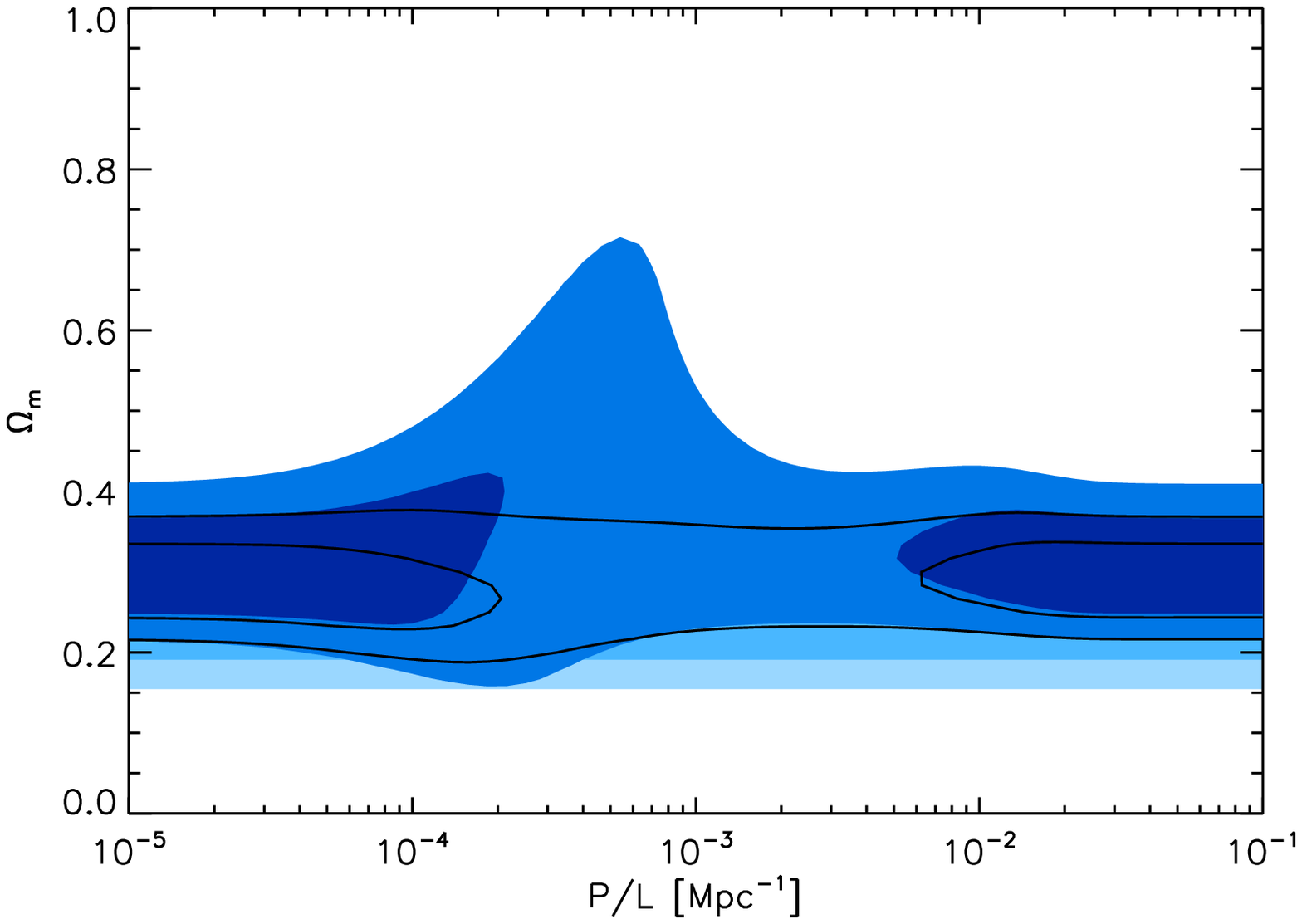}
    \includegraphics[height=2.8in,width=3in]{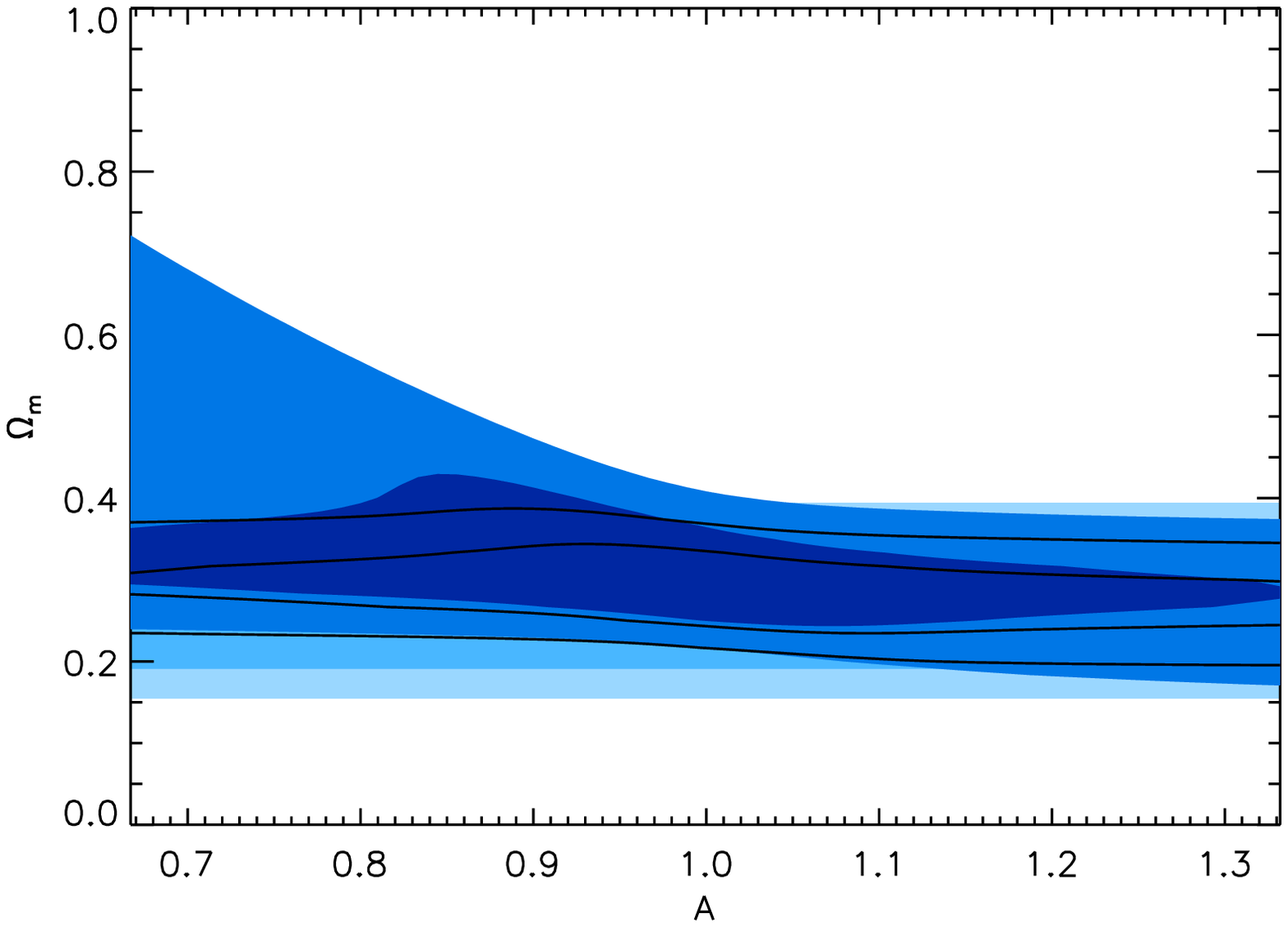} 
    \caption{\label{AOmm_POmm} 68\% and 95\% confidence levels (two parameters) 
                  on the $P/L-\Omega_m$ (left) and $A-\Omega_m$ (right) planes, for chameleons 
                  with a prior $A\in [2/3,4/3]$.  In the upper we have taken into account the order 
                  ${\mathcal O}(1)$ effects arising from the redshift dependance, while in the lowers
                   panels we have ignored these effects.  Dark blue contours are 
                  for SN data only, light blue ones for $H(z)$ data, and black 
                  transparent contours are for the joint SN$+H(z)$ analysis.}
   \end{center}
  \end{figure}

The first case is morphologically equivalent to the previous ALP case so 
the conclusions of the last section hold. 
In the  second case we see from Fig.~\ref{fig:redNored} that we
cannot constrain any value of the probability of conversion.
Our sensitivity is at most $|A-1|\gtrsim 0.1$ but only in a very narrow range 
of $P$ around $10^{-3} L/$Mpc.  Therefore, we cannot exclude the possibility 
that photons and chameleons mix strongly in the intergalactic medium, even 
though, in order to realise this scenario fully, more work is required to understand 
the possible fluxes of chameleons from  SNe.  For the third case we can obtain 
a constraint on the $P/L-\Omega_m$ plane by marginalizing over $A$ in the physically reasonable 
range $[2/3,4/3]$ (refer to Eq.~(\ref{Pnoredshift_and_A})).   Fig.~\ref{AOmm_POmm} shows our 
results in this case together with the corresponding constraint on the $A-\Omega_m$ plane 
after marginalising over $P/L$ in the range $[10^{-5},10^{-1}]$.  

Note that, in agreement with our analysis of the simplest ALP case with $A=2/3$, our constraints 
exclude these models as an alternative to a cosmological constant at greater than $3$-$\sigma$ if 
the Universe is set to be spatially flat, while $\Omega_m>0.8$ is excluded at $~2$-$\sigma$ even 
for SN data alone, and at much higher statistical significance  when $H(z)$ data are included. 
This is true both when the redshift dependence of the background is included  
(upper panels) or neglected  (lower panels), the constraint in the latter  case
being somewhat stronger. 

\section{\label{MCPs}Mini-Charged Particles/Hidden Photons}

New particles with a small unquantized charge have been investigated in several extensions of the standard model~\cite{Holdom:1985ag,Batell:2005wa}.
In particular, they arise naturally in extensions of the standard model which
contain at least one additional U(1) hidden sector gauge
group~\cite{Holdom:1985ag,Bruemmer:2009ky}. The gauge boson of this
additional U(1) is known as a hidden photon, and hidden sector
particles, charged under the hidden U(1), get an induced electric
charge proportional to the small mixing angle between the kinetic
terms of the two photons.
In string theory, such hidden U(1)s and the required kinetic mixing are a generic
feature~\cite{Abel:2006qt,Abel:2008ai,Dienes:1996zr,Abel:2003ue,Mark}.
Hidden photons are not {\it necessary} however to explain
mini-charged particles, and explicit brane-world scenarios have been
constructed \cite{Batell:2005wa} where MCPs arise without the need
for hidden photons.

The existence of low-mass MCPs can have a tremendous impact on 
photon propagation over cosmological distances. 
Photons from a given source can for instance pair produce MCPs with
CMB photons $\gamma+\gamma_{\rm CMB} \to \psi^++\psi^-$, leading to a
new form of opacity. However, this process is generally more
noticeable for CMB photons rather than those of higher energy, both because the CMB spectrum
was measured by the FIRAS experiment to be a perfect blackbody with a typical accuracy of $10^{-4}$, 
and, also, because the cross-section is inversely proportional to the center-of-mass energy. The impact of the existence of MCPs for CMB distortions was studied in~\cite{Melchiorri:2007sq}, where a limit for the minicharge $q_{\epsilon} < 4\times 10^{-8}$ (measured in units of the electron's charge) was derived for 4-component Dirac MCPs.   

A more relevant source of opacity was pointed out in \cite{Ahlers:2009kh}, following the work of \cite{Gies:2006ca,Ahlers:2007rd}. 
Photons propagating in a background magnetic field can actually pair-produce MCPs without the need for a second photon in the initial state.
This is due to the fact that in a background field energy-momentum conservation is non-trivial. 
Indeed, the magnetic field acts as a refractive medium where both the
mini-charged particles and photons, acquire a non-trivial dispersion relation, i.e. effective masses. 
In the most interesting case, the effective photon mass is larger than
that of a MCP pair, and the $\gamma\to \psi\bar \psi$ process happens at a rate
\begin{equation}
\kappa = \frac{e^{8/3} }{4 \Gamma\left(\frac{1}{6}\right)\Gamma\left(\frac{13}{6}\right)} \left(\frac{2B^{2}q_{\epsilon}^{8}}{3\omega}\right)^{1/3} f  \,,
\end{equation}
where $q_{\epsilon}$ is the MCP electric charge in units of the electron's charge $e$, and $f$ is an order one factor which depends on the nature of the MCP and the photon polarization with respect to the magnetic field, assumed again to be transverse to the photon direction of motion ($f=1,2/3$ for parallel and perpendicular polarizations respectively if the MCP is a Dirac spinor, and $f=1/12,1/4$ if the MCP has spin-0).  $\Gamma$ denotes the usual $\Gamma$-function.  The above formula is valid in the deep non-perturbative regime, where the adiabatic condition
\begin{equation}
\frac{3}{2}\frac{\omega}{m_\psi}\frac{eq_{\epsilon} B}{m_\psi^3} \ll 1
\label{adiabatic}
\end{equation}
holds (hence the unusual scaling with the charge, $(eq_{\epsilon})^{8/3}$). 
Note that in this regime the process is independent of the MCP vacuum
mass, $m_\psi$, but this parameter still  enters  through the adiabatic condition (\ref{adiabatic}). For the value of $\kappa$ in the non-adiabatic regime we refer to Appendix A of~\cite{Ahlers:2007rd}.

The MCP pair production process damps the photon flux according to the usual decay law, so the photon survival probability after traveling physical distance $L$ will be given by
\begin{equation}
{\cal P}(L) = \exp \left(- \int_0^L \kappa(L) dL \right)\,,
\end{equation}
where $L$ is redshift-dependent.  Using the redshift dependencies quoted in the 
discussion around Eq.~(\ref{redshiftdependence}), $\kappa$ redshifts\footnote{In the
non-adiabatic regime this scaling is not valid.} as $\kappa(1+z)$.  This leads us to 
\begin{equation}
{\cal P}(z) = \exp \left(- \kappa y(z)\right) \,, 
\end{equation}
where $y(z)$ is the comoving distance to the source.  Note that this expression can be recovered from 
the ALP case, Eq.~(\ref{weakprob}), in the $A\to 0$ limit and substituting $3P/(2L)\to \kappa$.

As was noted in~\cite{Ahlers:2009kh,Burrage:2009yz} the above expression does not hold in 
what is probably the most interesting case, in which the MCPs arise from kinetic mixing.
In that situation, photon to hidden photon oscillations also have to be
taken into account and, most surprisingly, they tend to suppress the
photon disappearance! In this scenario {\it both} photons and hidden
photons get an effective mass from the magnetic-field-dressed MCP. 
However, the coupling of hidden photons to  the MCP particles is much
stronger  than the corresponding coupling for photons, so the refractive effect 
(the effective mass) is always larger. 
The large mass of the hidden photon acts to suppress the mixing angle
between photons and MCPs, in a similar manner a large plasma frequency
(effective mass for the photon) suppresses the mixing between photons
and ALPs in the previous section.  The photon survival probability saturates at a value
\begin{equation}
{\cal P}(z) = 1-2\chi^2 \,,
\end{equation}
where $\chi$ is the kinetic mixing between photons and hidden photons.  
The interested reader can find further details about these arguments in~\cite{Burrage:2009yz}.
The value of $\chi$ is usually restricted to be smaller than $10^{-3}$
since it has a radiative origin, so there is no foreseeable constraint
on this scenario from cosmic opacity.  Therefore, we must focus our
attention on  the pure MCP scenario.

\begin{figure}[t]
  \begin{center}
    \includegraphics[height=2.8in,width=3in]{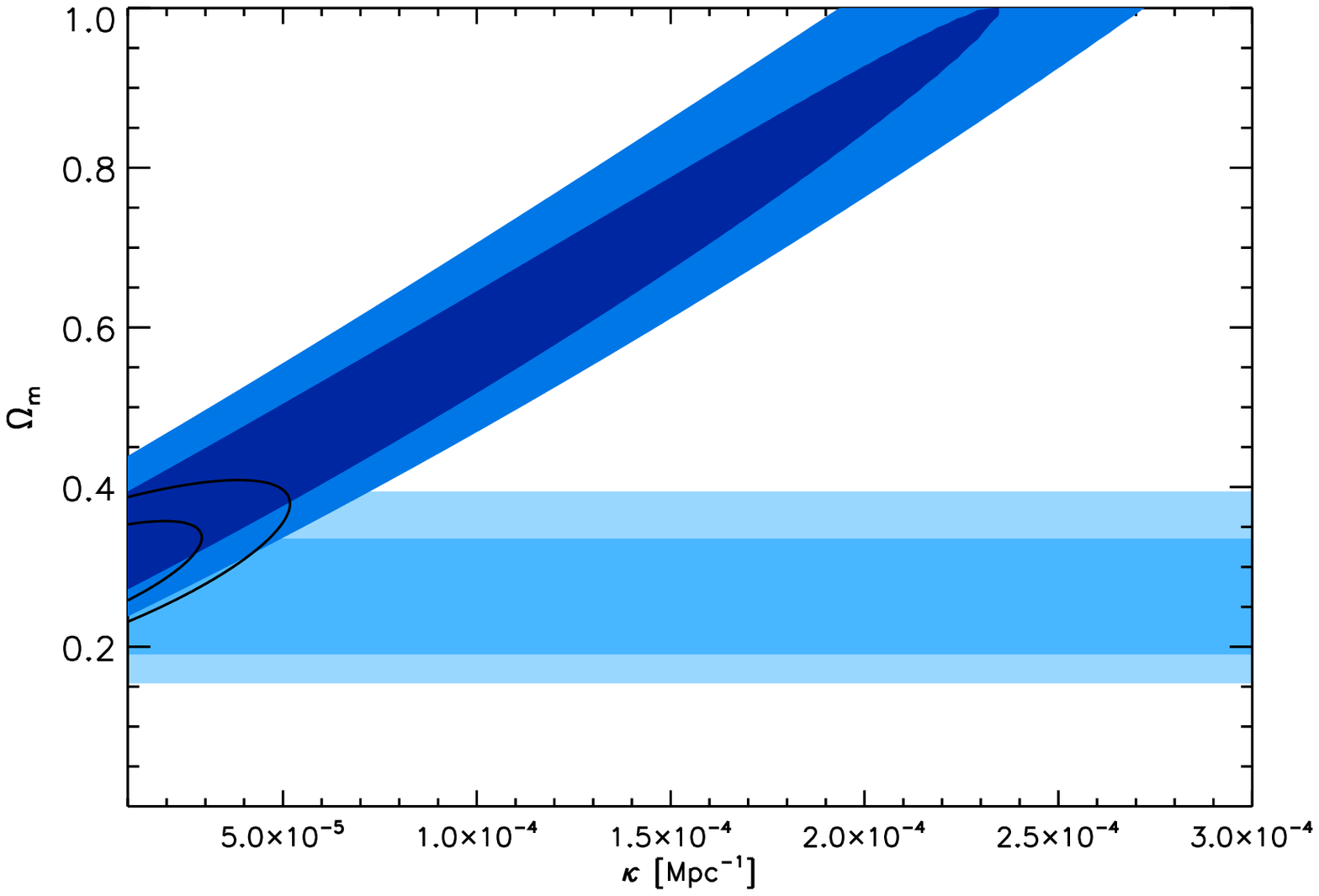}
    \includegraphics[height=2.8in,width=3in]{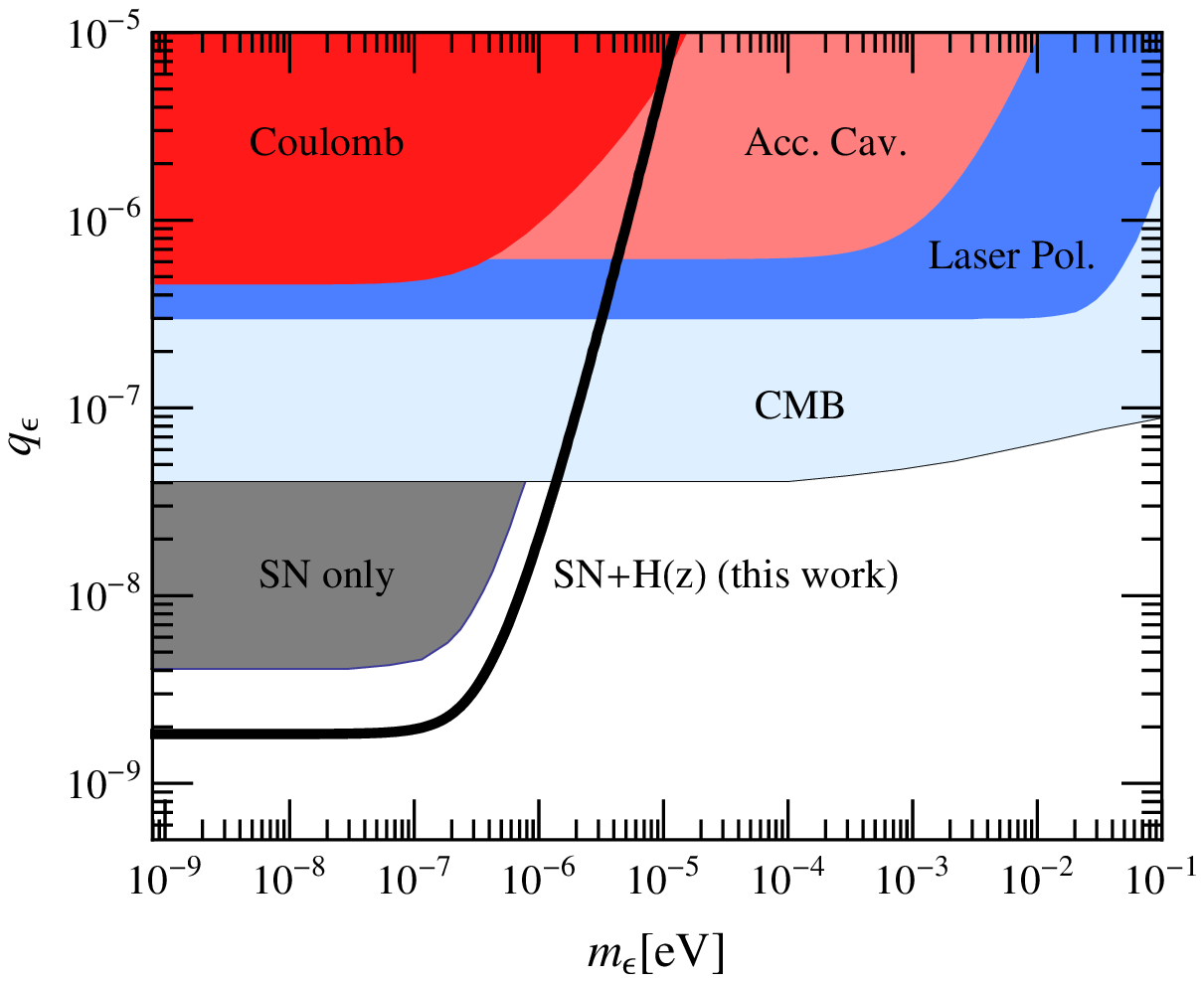} 
    \caption{\label{MCP_constrs} Confidence levels (68\% and 95\%) on the 
                  $\Omega_m-\kappa$ plane for the pure MCP model (left panel).
                  Dark blue contours show constraints from SN data only, light blue 
                  from $H(z)$ data, and solid line contours from joint SN$+H(z)$. 
                  The right panel shows the corresponding constraints in the mini-charge vs MCP mass plane assuming $B=1$nG. For comparison we show the previous SN dimming constraints~\cite{Ahlers:2009kh} (also for $B=1$nG), CMB bounds~\cite{Melchiorri:2007sq} and the most sensitive purely laboratory experiments, light polarization~\cite{Ahlers:2007qf}, tests of the Coulomb's law~\cite{Jaeckel:2009dh} and accelerator cavities~\cite{Gies:2006hv}.}
   \end{center}
  \end{figure}  

Fig.~\ref{MCP_constrs} (left) shows 1 and 2-$\sigma$ joint confidence levels in the $\kappa-\Omega_m$ 
plane, again for SN data only (dark blue contours) and for the combined SN+$H(z)$ data set (solid 
line contours).  In this case, SN data alone allow a zero cosmological constant in the presence of 
MCPs with a value $\kappa \sim 2.3\times 10^{-4}$ as suggested in~\cite{Ahlers:2009kh}. However, 
the inclusion of $H(z)$ data rules out this possibility and sets a strong bound 
\begin{equation}
\kappa < 5\times 10^{-5}\ \  {\rm Mpc}^{-1} \; (2-\sigma) \,,
\end{equation}
which, translated into MCP parameters, allows us to constrain the
region in parameter space shown in Fig.~\ref{MCP_constrs} (right).

\section{\label{forecasts}Forecasts for future baryon acoustic oscillations  and Supernovae surveys}

So far we have investigated constraints on cosmic opacity -- and also their 
implications for models which violate photon number conservation -- that are 
imposed from current data, namely from direct measurements of cosmic 
expansion $H(z)$ using  cosmic chronometers combined with Type 
Ia Supernova data (in particular the SCP Union 2008 compilation).  However, 
new and more accurate data for $H(z)$ (as well as $d_A(z)$, the angular 
diameter distance) will be available 
through ongoing and future  Baryon acoustic oscillations (BAO) surveys.  In this section, we show forecasted
constraints for cosmic opacity and the related models of sections 
\ref{cham}-\ref{MCPs}, that can be achieved by combining Supernova and  $H(z)$ data 
from future spectroscopic BAO surveys.  We focus in particular on two 
BAO missions, namely the Baryon Oscillation Spectroscopic Survey (BOSS)
and EUCLID.  Finally, we also consider forecast constraints from proposed 
SN missions, in particular combining EUCLID and SNAP. 

\begin{figure}[h]
  \begin{center}
    \includegraphics[height=4in,width=4.1in]{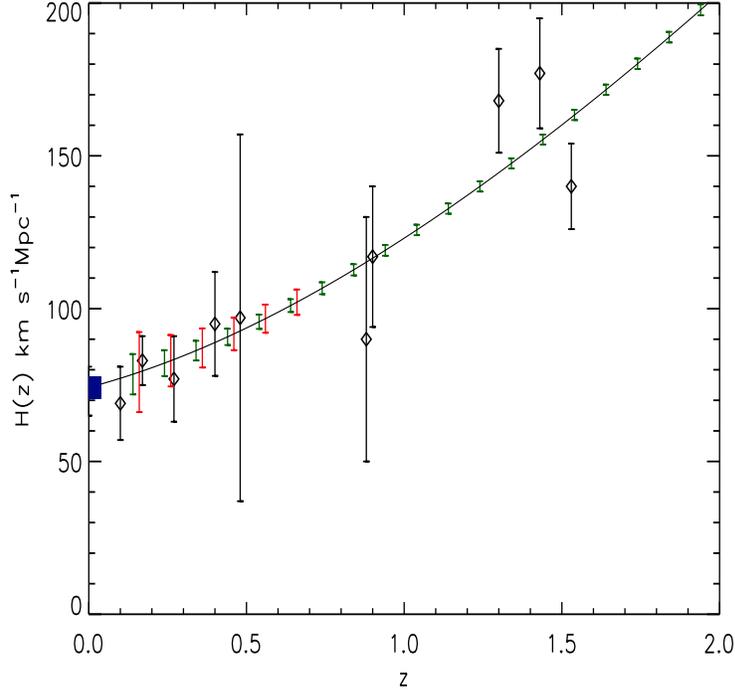} 
    \caption{\label{forecast_errs} Forecasted error-bars on $H(z)$ for BOSS (red) 
                  and EUCLID (green) compared to current $H(z)$ data from differential 
                  ageing of galaxies (black).  Also shown in the blue rectangle is the 
                  Riess et al. determination of $H_0$ \cite{Riess09}.}
  \end{center}
\end{figure}

BOSS \cite{BOSS} is part of the SDSS-III survey and is scheduled 
to operate over the period 2009-2014.  Using the 2.5 m SDSS telescope, it 
will measure redshifts of 1.5 million luminous galaxies in the range $0.1<z<0.7$
(as well as Ly$\alpha$ absorption towards 160,000 high-redshift quasars at about 
$z\simeq 2.5$), covering $\simeq$10,000 ${\rm deg}^2$ of high-latitude sky.  The forecast 
precision for $H(z)$ is $1.8\%$, $1.7\%$ and $1.2\%$ in  redshifts bins centered at  $z=0.35$, 0.6 and 
2.5 respectively.  On the other hand, EUCLID --  proposed to ESA's Cosmic Visions
programme -- aims for lunch around 2018.  A combination of the earlier 
SPACE \cite{SPACE} and DUNE \cite{DUNE} missions, 
EUCLID would provide around 150 million redshifts in the range $z<2$, covering 
about 30,000 ${\rm deg}^2$.  Fig. \ref{forecast_errs} shows forecast errors around
the WMAP7 $\Lambda CDM$ model for both BOSS and EUCLID.  Also shown for 
comparison are the current $H(z)$ ``cosmic chronometers'' data used above.

\begin{figure}[h]
  \begin{center}
    \includegraphics[height=5in,width=5.12in]{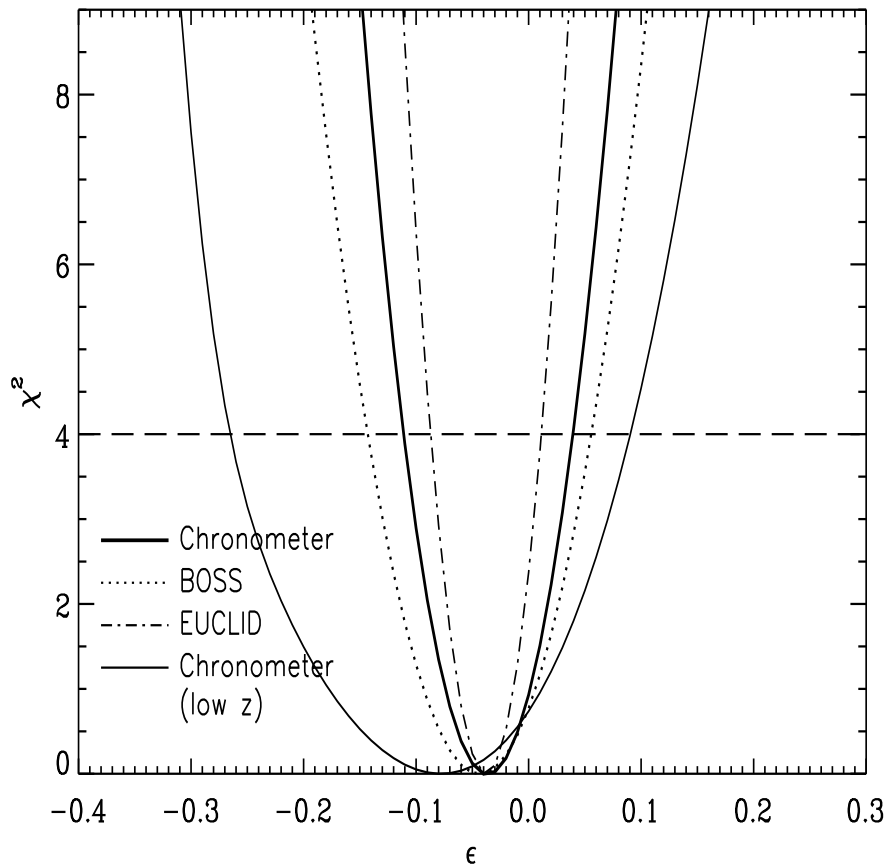} 
    \caption{\label{forecast_eps} Forecasted constraints on the opacity parameter 
                  $\epsilon$ of section \ref{update} for BOSS and EUCLID, combined with 
                  current SN data.  Also shown are current constraints from $H(z)$ 
                  ``chronometer'' data (again joint with SN). The small shift towards negative $\epsilon$ is due to the fact that actual SN data are being used.}
  \end{center}
\end{figure}

We use the code developed by Seo \& Eisenstein \cite{SeoEisen} to estimate
the errors in radial distances achievable by using BAOs as a standard 
ruler.  Fig. \ref{forecast_eps} shows our forecasted constraints on the parameter 
$\epsilon$ of section \ref{update}, using the current type Ia SN data (Union 2008)
in combination with modelled BAO data with forecasted errors for both BOSS and  
EUCLID\footnote{We will consider forecasted constraints combining planned SN 
missions' data  as well at the end of this section.}.  Note that although BOSS 
will achieve much smaller error bars than those of current $H(z)$ data 
(cf Fig. \ref{forecast_errs}), it will span a much narrower redshift range, so it will 
in fact provide somewhat weaker constraints than the current $H(z)$ 
``chronometers'' data.  To make a more direct comparison  we have also shown 
the corresponding constraints obtained by restricting the current $H(z)$ data in 
the narrower redshift range available to BOSS (thin solid line labeled ``chronometer 
(low z)''). On the other hand, significant improvement of these constraints will be 
achieved by EUCLID.  

\begin{figure}[h]
  \begin{center}
    \includegraphics[height=2.8in,width=3in]{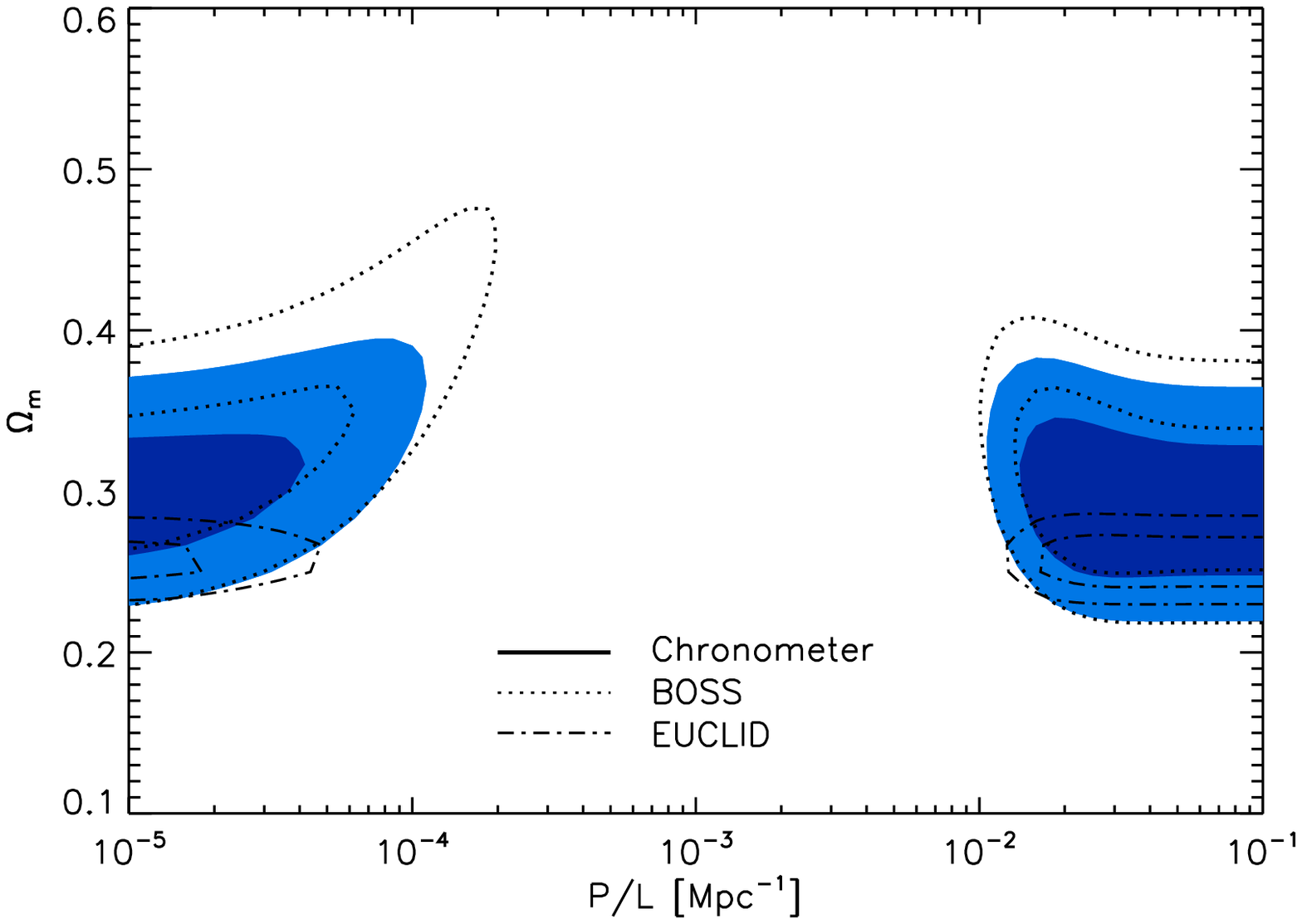}
    \caption{\label{forecast_ALP} The joint constraints of Fig.~\ref{POmm} 
                  for the simple ALP model of section \ref{cham} with $A=2/3$, shown 
                  together with the corresponding forecast constraints from BOSS and 
                  EUCLID combined  with current SN data.}
  \end{center}
\end{figure}

\begin{figure}[h]
  \begin{center}
    \includegraphics[height=2.8in,width=3in]{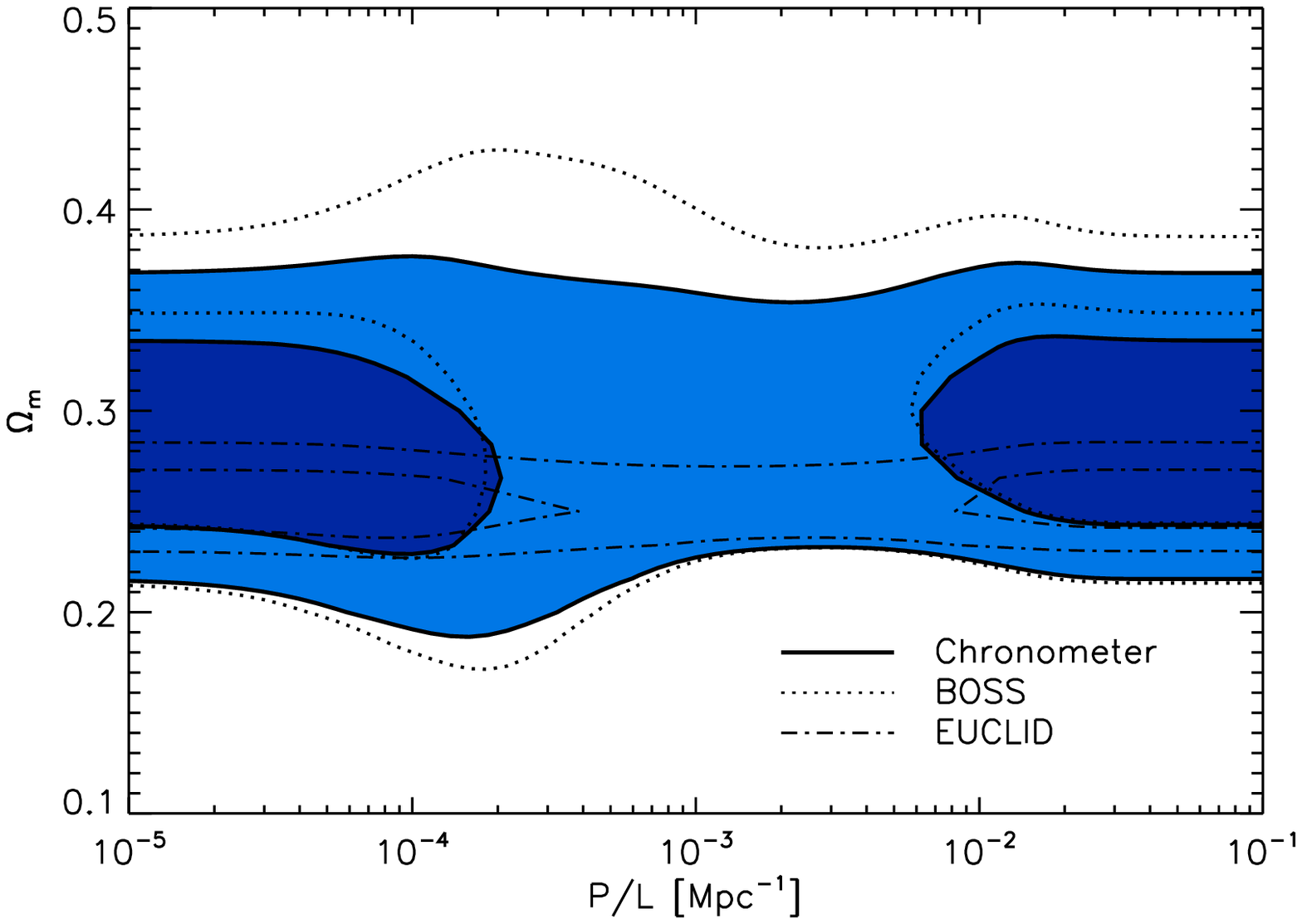}
    \includegraphics[height=2.8in,width=3in]{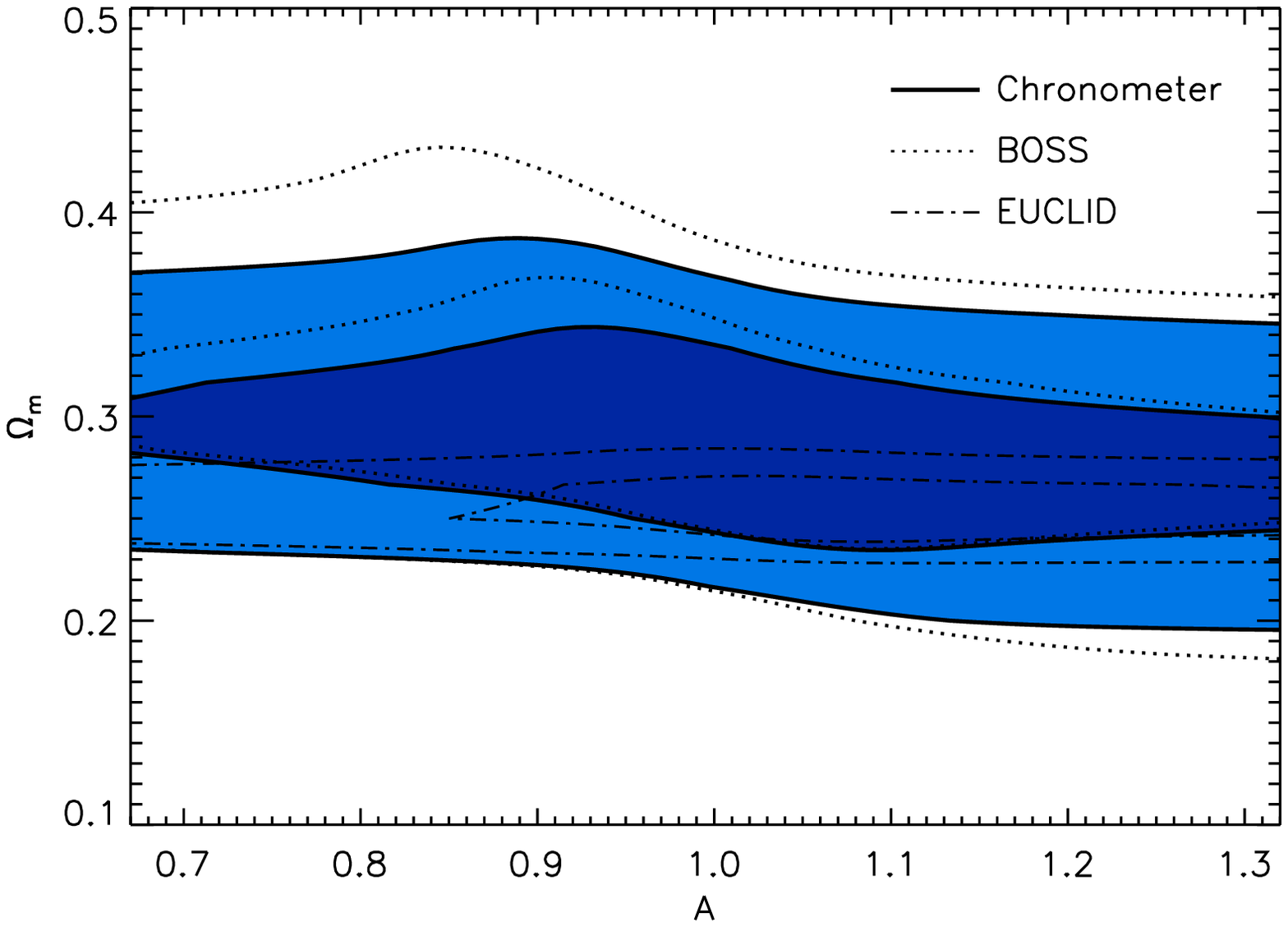}  
    \caption{\label{forecast_cham} The joint constraints of Fig.~\ref{AOmm_POmm} for 
                  chameleons with $2/3<A<4/3$), together with the corresponding forecast 
                  constraints from BOSS and EUCLID joint with current SN data.}
  \end{center}
\end{figure}

\begin{figure}[h]
  \begin{center}
    \includegraphics[height=2.8in,width=3in]{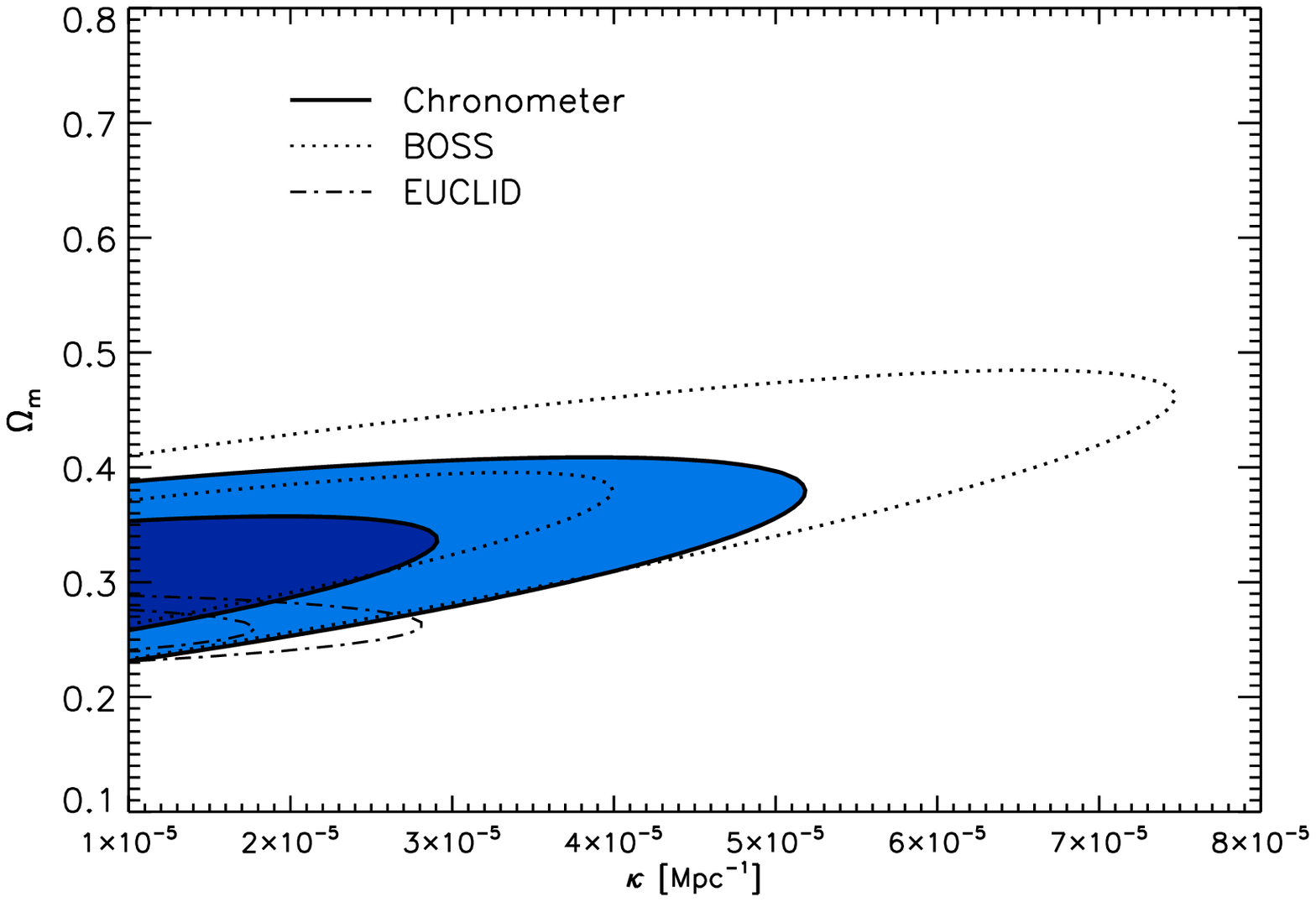}
    \caption{\label{forecast_MCPs} Constraints of Fig. \ref{MCP_constrs}
                  for MCPs, compared to forecasted constraints from BOSS and 
                  EUCLID, combined with current SN data.}
  \end{center}
\end{figure}

Similarly, in Figs.~\ref{forecast_ALP} \& \ref{forecast_cham} we show forecasted
constraints for the simple ALP model of section \ref{cham} ($A=2/3$) and for 
chameleons with $A\in[2/3,4/3]$ (section \ref{cham}).  For ALPs, EUCLID 
will provide significant improvement, notably by a factor of 2-3 on the constraints 
on $P/L$ in the weak mixing regime of Fig.~\ref{forecast_ALP}.  As the probability 
of mixing is inversely proportional to the square of the energy scale of the ALP 
coupling, this will result in a modest improvement of the bounds on M by factor of 
order unity.   Finally, forecast constraints for MCPs (section \ref{MCPs}) are displayed 
in Fig. \ref{forecast_MCPs}, where, again, EUCLID will improve constraints on the 
parameter $\kappa$ by a factor of 2-3.  As $\kappa \sim q_{\epsilon}^{8/3}$, this 
results in a modest, order unity improvement of the constraints on the charge of 
any exotic MCP.  

\begin{figure}[h]
  \begin{center}
    \includegraphics[height=1.9in,width=2in]{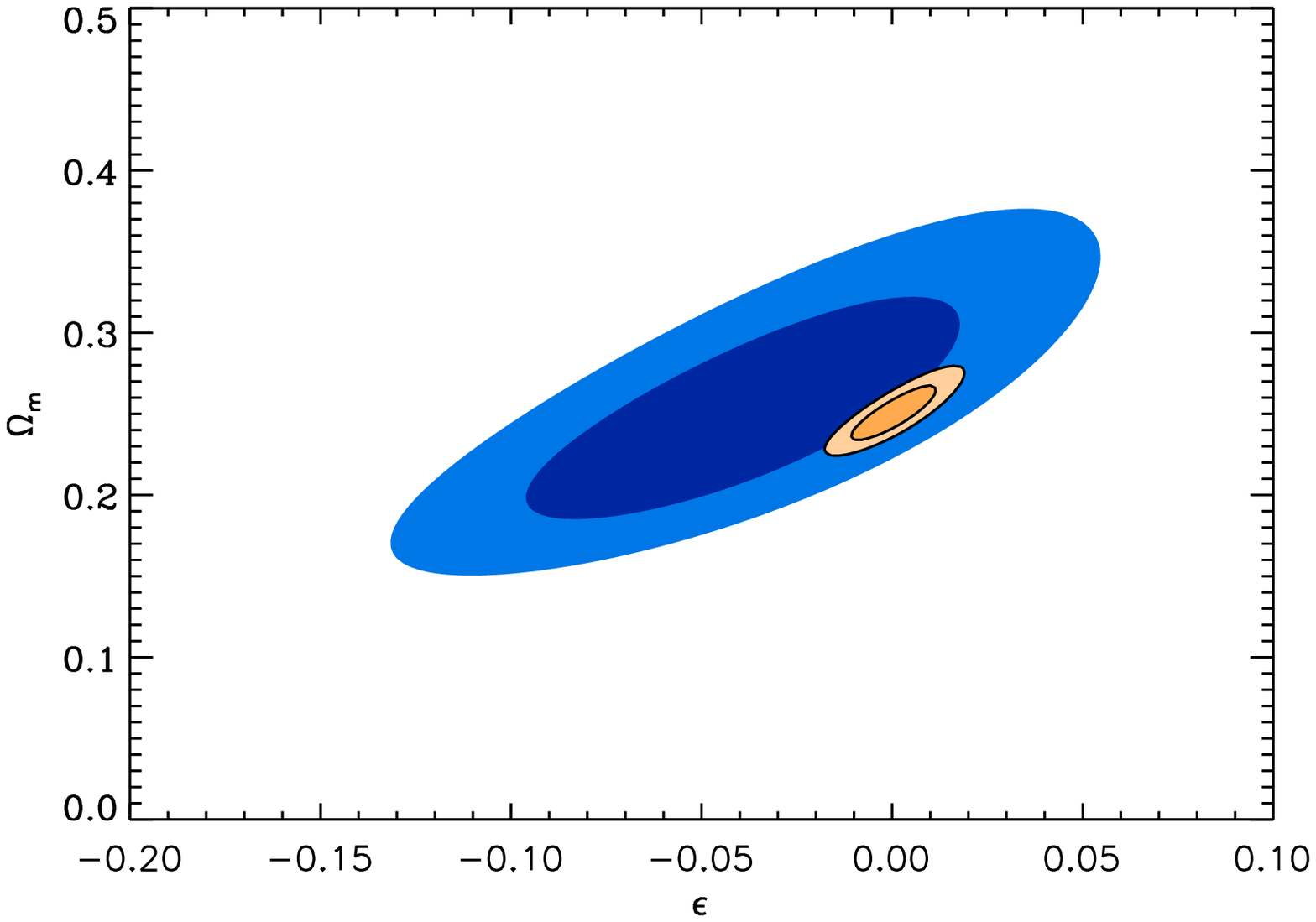} 
    \includegraphics[height=1.9in,width=2in]{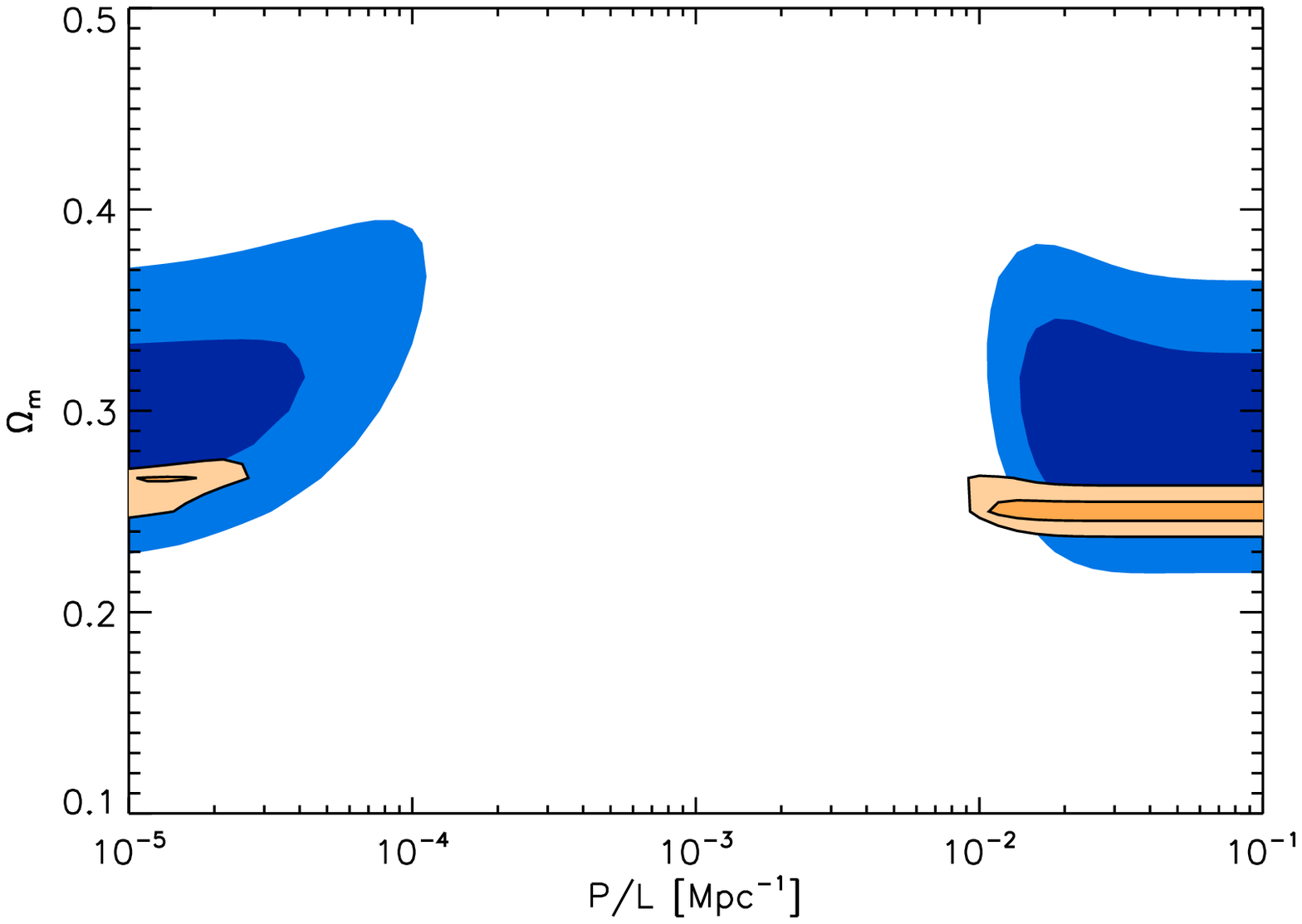}\\
    \includegraphics[height=1.9in,width=2in]{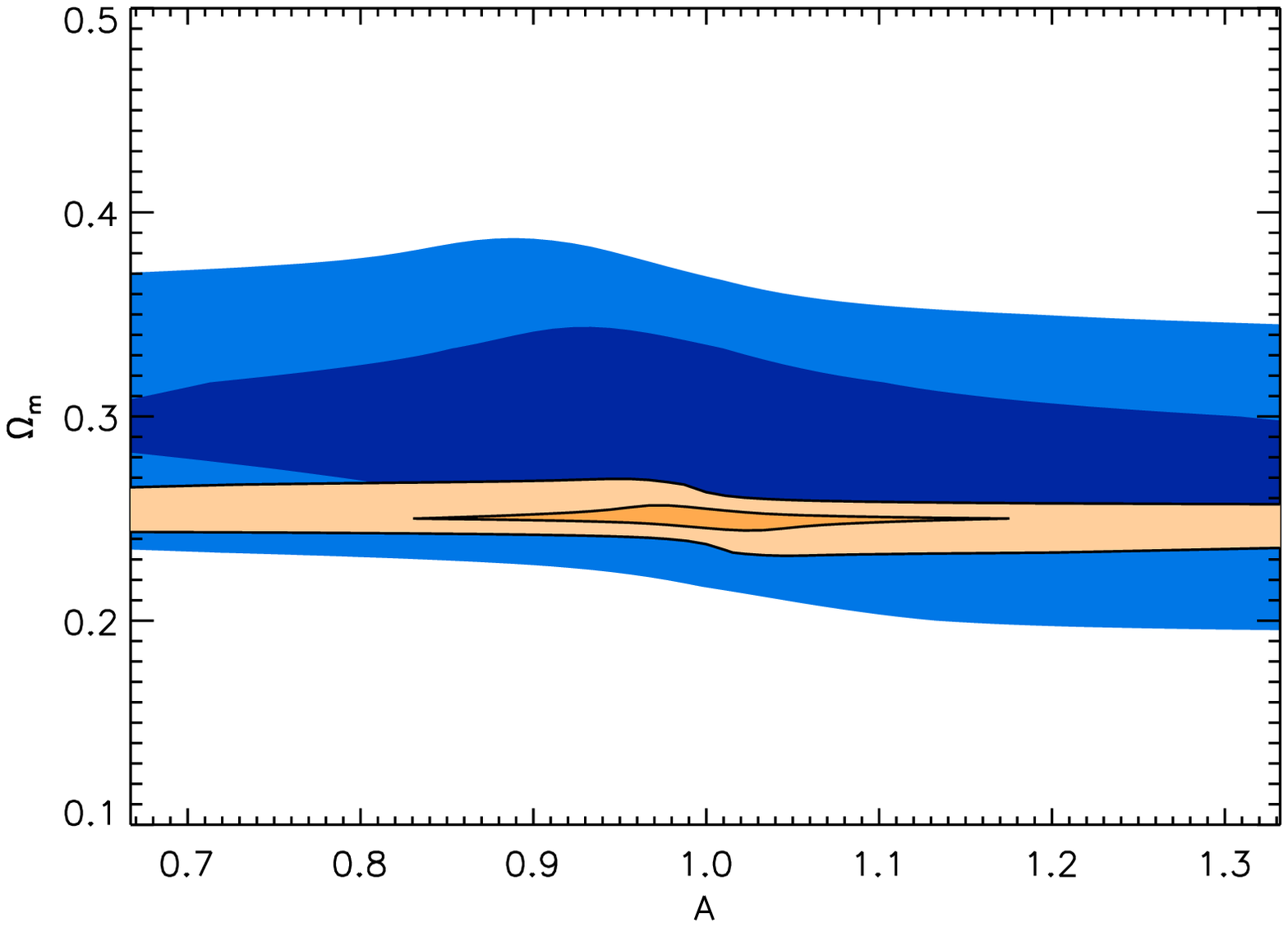}
    \includegraphics[height=1.9in,width=2in]{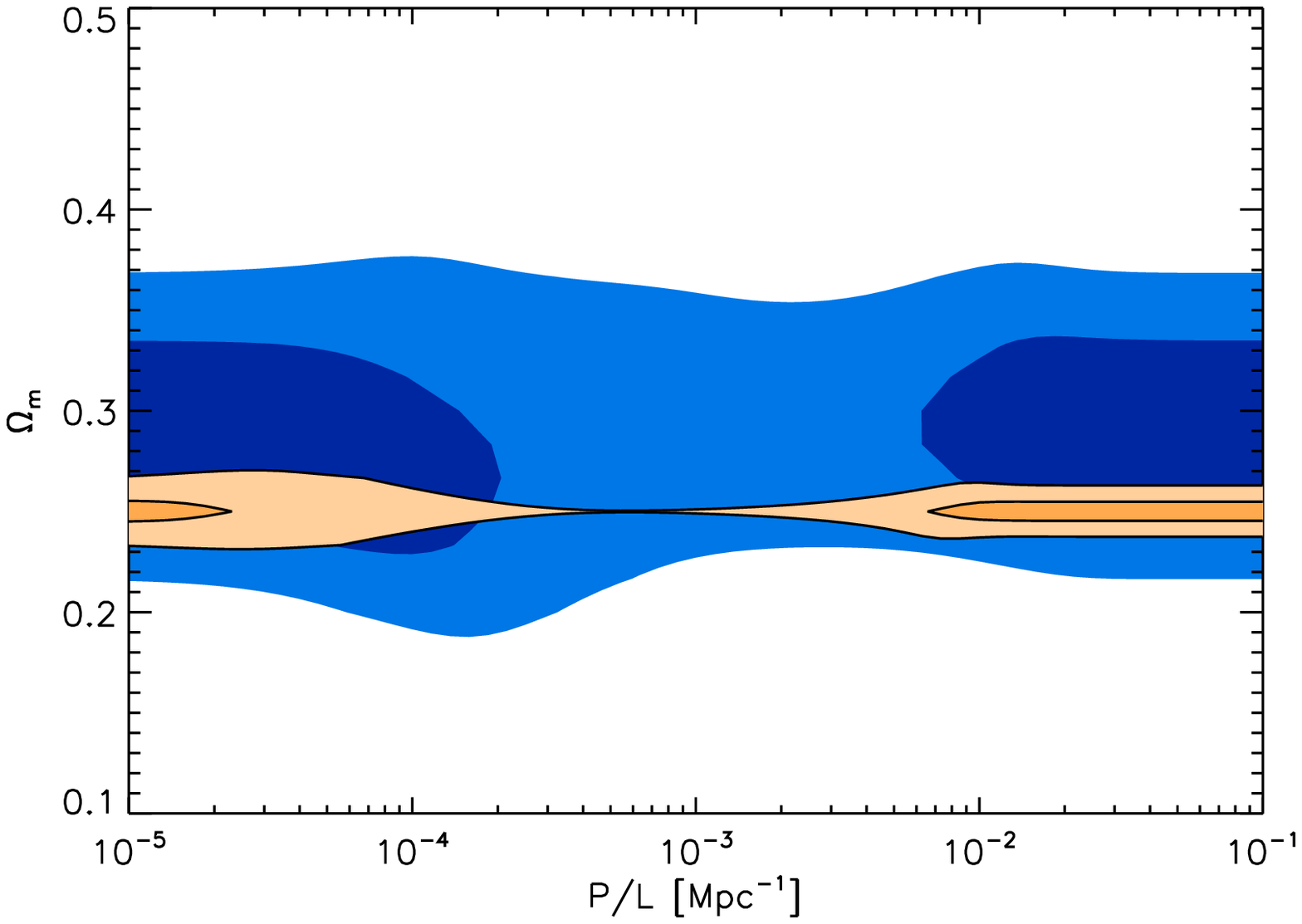}
    \includegraphics[height=1.9in,width=2in]{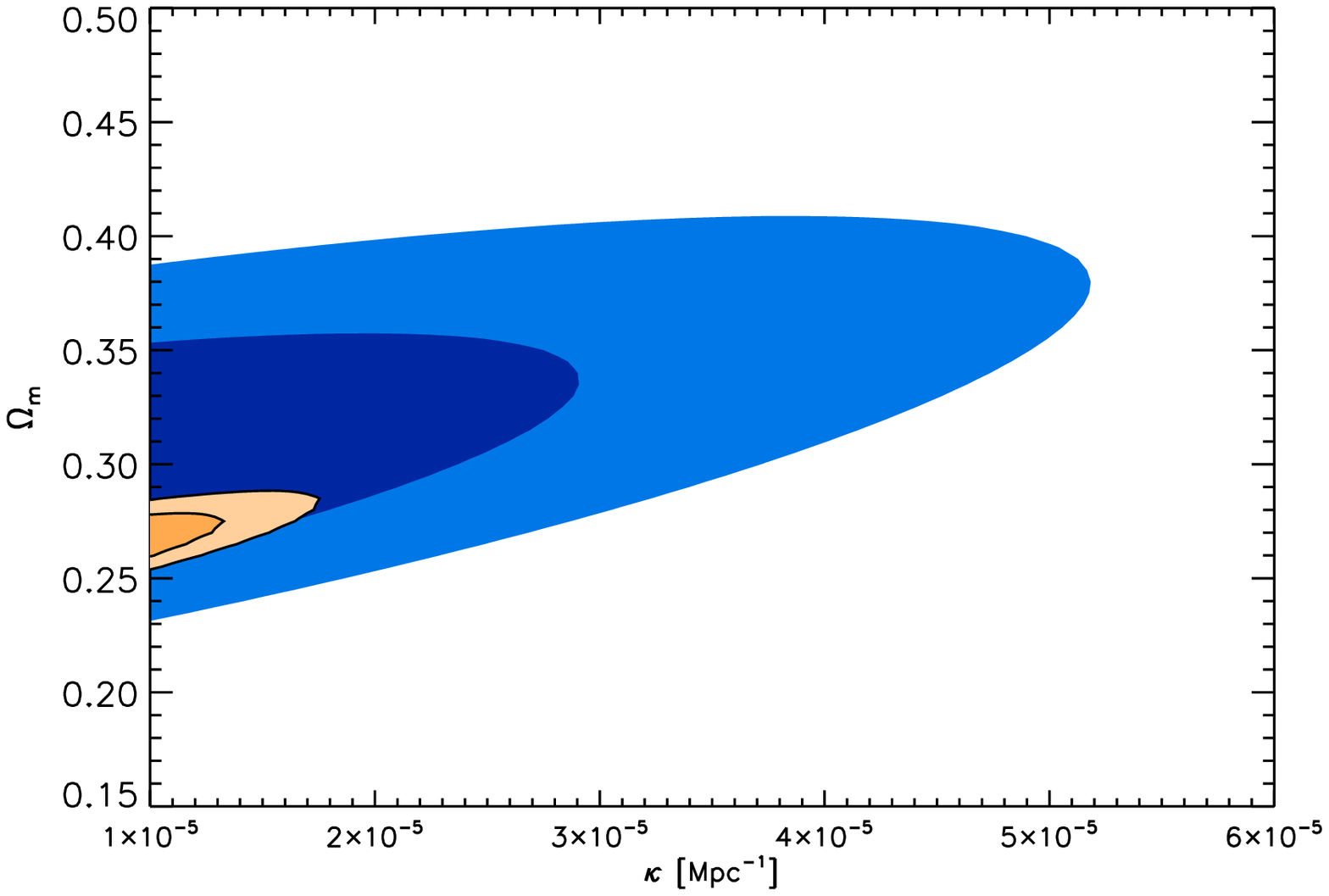} 
    \caption{\label{SNBAO_forec} Forecast constraints from joint SNAP+EUCLID 
                   (orange scale), shown together with the corresponding constraints from 
                   current data, namely SN (Union08) joint with chronometer $H(z)$ (blue 
                   scale).  Dark and light contours correspond to 1- and 2-$\sigma$ respectively.  
                   From top left to bottom right: constraints on the opacity parameter $\epsilon$ 
                   (section \ref{update}), parameter $P/L$ for the simple ALP model of section 
                   \ref{cham}, parameters $A$ \& $P/L$ for chameleons (section \ref{cham}), 
                   and parameter $\kappa$ for MCPs (section \ref{MCPs}).}
  \end{center}
\end{figure}

So far we have considered the effect that future BAO data will have on 
the constraints of sections \ref{update}-\ref{MCPs}, when combined with 
current SN data, and in particular the SCP Union 2008 compilation.  We 
showed that EUCLID will lead to a significant improvement of these 
constraints (figures \ref{forecast_eps}-\ref{forecast_MCPs}), while the 
narrower redshift range of BOSS renders it comparable to current $H(z)$ 
measurements for constraining these models.  We end this section by   
considering the effect that proposed SN surveys data will have on these constraints, 
in particular forecasted constraints for SNAP (or dark energy task force stage 
IV SNe mission) \cite{DETF} combined with EUCLID.  Fig. \ref{SNBAO_forec} summarises 
these constraints for all models considered above.  Our forecasted constraints, 
shown in orange scale, appear on top of the corresponding joint SN+$H(z)$ 
constraints from current data, see sections 3 and 4.  SN data from these proposed surveys will lead to notable improvement of these constraints, 
for example, by nearly an order of magnitude in the parameter $\epsilon$
described above.  As the figure shows, for the models considered above, this will correspond to 
an improvement of a factor of up to few on the strength of the coupling 
of ALPs to photons, and on the charge of MCPs.    

\section{Conclusions}

If new particles from physics beyond the standard model couple to
photons then the propagation of light may be altered.  In this paper
we have focused on two scenarios for exotic particles  which can
significantly modify the propagation of photons as they pass through
magnetic fields.    Measurements of cosmic opacity are a strong tool to constrain
such scenarios, as interactions between photons and exotic particles
in the magnetic fields of the intergalactic medium leads to a new
source of cosmic opacity.
Uniform deviations from cosmic transparency (i.e. opacity) can be constrained through their 
effects on distance duality, by parameterizing possible deviations from the Etherington relation.
The Etherington relation  implies that, in a cosmology based on a metric theory 
of gravity, distance measures are unique: the luminosity distance is $(1 + z)^2$ times the 
angular diameter distance. Both luminosity distance and angular diameter distance depend on the Hubble parameter $H(z)$, but this relation is valid in any cosmological background where photons 
travel on null geodesics and where, crucially, photon number is conserved. We have   restricted our attention on violations of the Etherington relation arising from the  violation of photon conservation.

We have combined direct measurements of cosmic expansion (from the latest determinations of the Hubble parameter)  at redshifts $0<z<2$ and recent SN data yielding the luminosity distance.  SN-inferred luminosity distances are affected by violation of photon conservation, but the $H(z)$ measurements  we use are not.
Assuming an underlying flat $\Lambda$CDM model, we have placed tight limits on possible deviations from photon-conservation.
Photon-conservation can be violated by simple astrophysical effects which give uniform attenuation such as gray dust. We have reported updated constraints on this effect.

More exotic sources of photon-conservation violation involve a coupling of photons 
to particles beyond the standard model of particle physics. We have focused on axion-like particles, new scalar or pseudo scalar
fields which couple to the kinetic terms of photons, and
mini-charged particles which are hidden sector particles with  a tiny electric charge.  Photons
passing through intergalactic magnetic fields may be lost by pair
production of light
mini-charged particles.  If the mixing between axion-like particles
and photons is significant, then interactions in the intergalactic magnetic
fields will also lead to a loss of photons due to conversion into
ALPs.  However if the coupling between photons and ALPs is sufficiently
strong, one-third of any initial flux will be converted into ALPs,
and two-thirds into photons, resulting in a redshift-independent
dimming of supernovae which we cannot constrain or exclude with
cosmic opacity bounds. 

\begin{figure}[h]
  \begin{center}
    \includegraphics[width=3in]{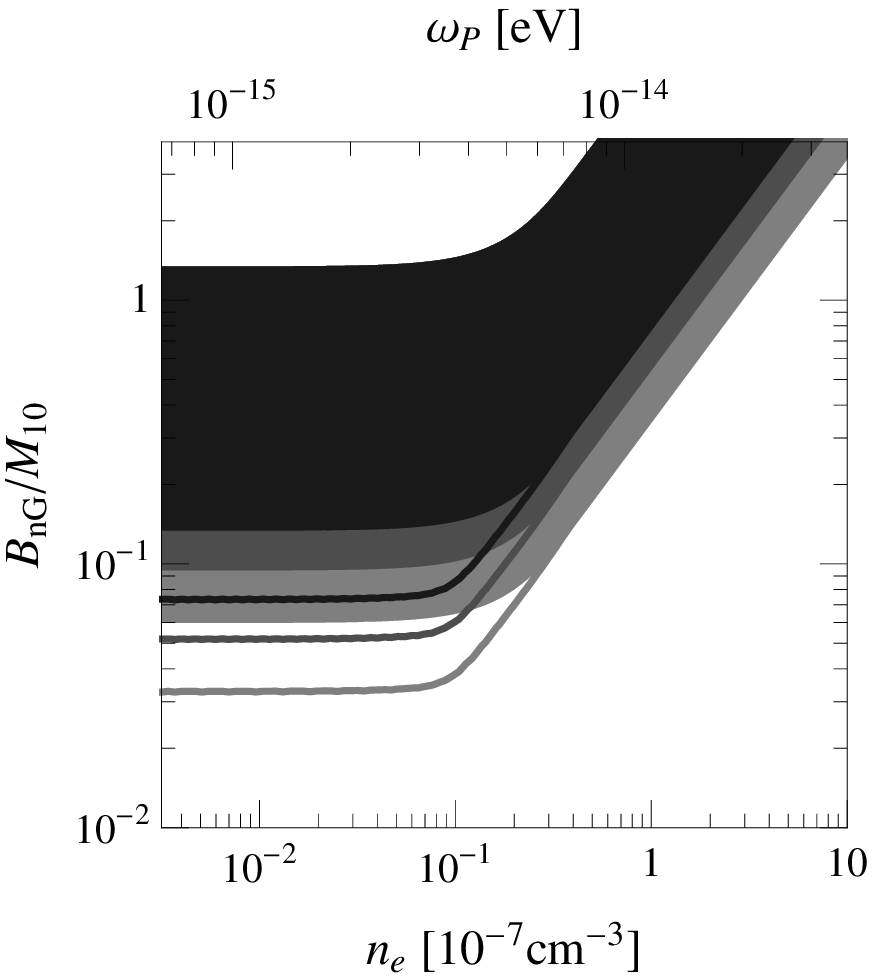} 
    \includegraphics[height=3in,width=3in]{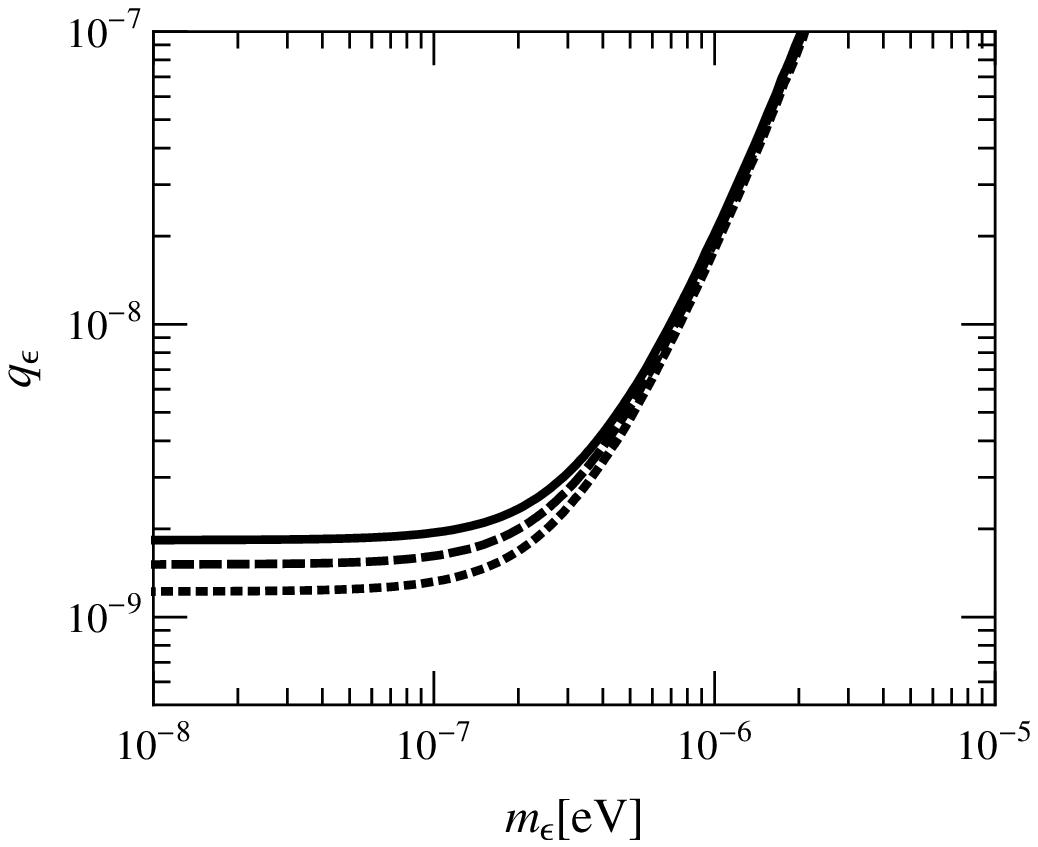} 
    \caption{\label{sumario} Prospects of the future bounds on ALPs (left) and MCPs (right).
    ALPs: The dark region corresponds to the current bounds while gray and light grey regions cover the possible future bounds using EUCLID and EUCLID+SNAP data.
    Taking into account the redshift dependence of the background improves these bounds in the region covered by the lines.
    MCPs: Current bounds (solid) and prospects from EUCLID (dashed) and EUCLID+SNAP (dotted).}
  \end{center}
\end{figure}

The improved measurement of the cosmic opacity found here leads to
improved bounds on these exotic physics scenarios which are summarised
in Fig. \ref{sumario}.  Future measurements of baryon acoustic
oscillations, and an increase in the number of observations of high
redshift supernovae will lead to further improvements in the
constraints on physics beyond the standard model. 

\section*{Acknowledgements}
The work of AA is funded by a CTC Postdoctoral Fellowship.  AA would like to thank the Institute of Cosmos Sciences, University of Barcelona, for hospitality and support.  CB is supported by the German Science Foundation (DFG) under the Collaborative Research Centre (SFB) 676.  JR acknowledges support from the DFG cluster of excellence EXC 153 Origin and Structure of the Universe.  LV is supported by FP7-PEOPLE-2007-4-3 IRG n 202182 and FP7 IDEAS-Phys.LSS.240117.  RJ is supported by FP7-PEOPLE-2007-4-3 IRG.  LV and RJ are also supported by MICINN grant AYA2008-03531.

\section*{Appendix}
Here we summarise the relevant equations describing how the luminosity distance measures are 
affected by the presence of particles beyond the standard model that couple to photons:  axion-like 
particles (ALP), chameleons and mini-charged particles (MCP).

The presence of ALPs will have an impact on observations of supernovae if they are 
observed through magnetic fields.  The strength and spatial structure of the inter-galactic 
magnetic fields is highly unconstrained and so they are left as parameters.  The luminosity 
distance to SNe is modified by the redshift-dependent probability of photon survival 
${\cal P}(z)$:
\begin{equation}
d_L(z)\rightarrow d_L(z)/\sqrt{{\cal P}(z)}\,.
\end{equation} 

There are two physical limits in which ${\cal P}(z)$ can be analytically calculated, 
namely the \emph{incoherent} and \emph{coherent} regimes.  In the incoherent 
regime one has:
\begin{equation}
{\cal P}(z)= A + (1-A) \exp\left(-\frac{3}{2}\frac{y(z)}{L}P\right) \,,
\end{equation}
where $L$ is the size of magnetic domains, $P$ the probability of 
photon to ALP conversion, and $y(z)$ the comoving distance to the
source.  In the coherent regime, one can take into account additional effects 
due to evolution of the background magnetic fields with redshift.  In this case 
the probability of photon survival reads: 
\begin{equation}
{\cal P}(z)= A + (1-A) \exp\left(-\frac{P}{H_0 L }\frac{H(z)-H_0}{\Omega_m H_0}\right) \,,  
\end{equation}
where $H_0$ and $\Omega_m$ are the Hubble constant and matter density 
parameter (at the present epoch) respectively, and $H(z)$ is the Hubble parameter 
at redshift $z$.    
For the simplest ALP models, $A=2/3$ in the above equations, while for chameleons 
these equations hold with $A$ $\ne 2/3$.

The existence of low-mass  MCPs can also have a tremendous impact on photon propagation 
over cosmological distances.  Photons from a given source can for instance pair produce 
MCPs, even without the need for a (CMB) photon, in the presence of an inter-galactic magnetic 
field.  The MCP pair production process damps the photon flux, so, again, the luminosity 
distance to SNe is modified by a redshift-dependent probability of photon survival, ${\cal P}(z)$.
In this case one has:
\begin{equation}
{\cal P}(z) = \exp \left(- \kappa y(z)\right) \,, 
\end{equation}
where $\kappa$ is the photon to MCP transition rate and is given by
\begin{equation}
\kappa = \frac{e^{8/3} }{4 \Gamma\left(\frac{1}{6}\right)\Gamma\left(\frac{13}{6}\right)} \left(\frac{2B^{2}q_{\epsilon}^{8}}{3\omega}\right)^{1/3} f \,.
\end{equation}
Here, $B$ denotes the magnetic field strength, $\omega$ is the photon energy, $q_{\epsilon}$ is the MCP electric charge in units of the electron charge $e$, and $f$ is an order unity factor which depends on the nature of the MCP and the photon polarization with respect to the magnetic field. $\Gamma$ denotes the usual $\Gamma$-function.


\section*{References}

\begin{thebibliography}{}

\bibitem{Etherington1} Etherington, J. M. H. 1933, Phil. Mag., 15, 761
\bibitem{Menard} M{\'e}nard, B.,
Nestor, D., Turnshek, D., Quider, A., Richards, G., Chelouche, D.,
\& Rao, S.  2008, MNRAS, 385, 1053
\bibitem{Bovy} Bovy, J., Hogg, D.~W.,
\& Moustakas, J. 2008, ApJ, 688, 198
\bibitem{Aguirre} Aguirre, A.~N. 1999, ApJL,
512, L19
\bibitem{Jaeckel:2010ni}
  Jaeckel, J., \& Ringwald, A., arXiv:1002.0329 [hep-ph]
\bibitem{Csaki2} Cs{\'a}ki, C.,
Kaloper, N., \& Terning, J., Phys. Rev. Lett. {\bf 88} (2002) 161302 
  \bibitem{Mortsellaxions} M{\"o}rtsell, E., 
Bergstr{\"o}m, L., \& Goobar, A., Phys.\ Rev.\  D {\bf 66} (2002) 047702 
\bibitem{Mirizzi:2006zy}
 Mirizzi, A., Raffelt, G.~G., \& Serpico, P.~D.,
 Lect.\ Notes Phys.\  {\bf 741}, 115 (2008)
 [arXiv:astro-ph/0607415]
\bibitem{Burrage:2007ew}
  Burrage, C.,
   Phys.\ Rev.\  D {\bf 77} (2008) 043009
  [arXiv:0711.2966 [astro-ph]]
\bibitem{Ahlers:2009kh}
  Ahlers, M.,
  Phys.\ Rev.\  D {\bf 80} (2009) 023513
  [arXiv:0904.0998 [hep-ph]]
 \bibitem{bassettkunz1} Bassett B. A., \& Kunz, M.,  Astrophys. J. {\bf 607}, 661 (2004)
\bibitem{bassettkunz2}   Bassett B. A., \& Kunz, M., Phys. Rev. D {\bf 69}, 101305 (2004)
\bibitem{Song:2005af}
Song,  Y.~S., \& Hu, W., 
 Phys.\ Rev.\  D {\bf 73} (2006) 023003
 [arXiv:astro-ph/0508002]
\bibitem{More} More, S., Bovy, J., \& Hogg, D.~W. 2008, arXiv:0810.5553
\bibitem{AVJ}
  Avgoustidis, A., Verde, L.,  \& Jimenez, R.,
  JCAP {\bf 0906} (2009) 012
  [arXiv:0902.2006 [astro-ph.CO]]
\bibitem{Union} Kowalski {\it et al.} 2008, ApJ, 686, 749
\bibitem{JVST} Jimenez, R., Verde, L.,
Treu, T., \& Stern, D.,\ 2003, ApJ, 593, 622
\bibitem{SVJ05} Simon J., {\it et al.}, Phys. Rev. D {\bf 71} (2005) 123001
\bibitem{Kostelecky:2008ts}
  Kostelecky, V.~A., \& Russell, N. 2008,
  arXiv:0801.0287 [hep-ph]
\bibitem{Uzan}
  Uzan, J.~P. 
  Gen.\ Rel.\ Grav.\  {\bf 39} (2007) 307
  [arXiv:astro-ph/0605313]
\bibitem{LazNesPer}
   Lazkoz, R., Nesseris, S. \& Perivolaropoulos, L. 2008, 
  JCAP {\bf 0807} 012
   [arXiv:0712.1232 [astro-ph]]

\bibitem{PercivalBAO} Percival, W.~J., Cole,
S., Eisenstein, D.~J., Nichol, R.~C., Peacock, J.~A., Pope, A.~C.,
\& Szalay, A.~S.  2007, MNRAS, 381, 1053
\bibitem{SJVKS}
  Stern, D., Jimenez, R., Verde, L., Kamionkowski, M., \& Stanford, S.~A., arXiv:0907.3149 [astro-ph.CO]
\bibitem{Riess09}
  Riess, A.G., {\it et al.},
  Astrophys.\ J.\  {\bf 699} (2009) 539
  [arXiv:0905.0695 [astro-ph.CO]]
\bibitem{HSTKey} Freedman, W.~L., {\it et al.} 2001, ApJ, 553, 47
\bibitem{kessler} Kessler, R., {\it et al.} 2009, ApJS, 185, 32   
\bibitem{Peccei:1977hh}
  Peccei, R.~D., \& Quinn, H.~R.,
  Phys.\ Rev.\ Lett.\  {\bf 38} (1977) 1440.
\bibitem{Svrcek:2006yi}
  Svrcek, P., \& Witten, E.,
  JHEP {\bf 0606} (2006) 051
  [arXiv:hep-th/0605206]  
\bibitem{Anselm:1981aw}
	Anselm, A.~A., \& Uraltsev, N.~G.,
	Phys. Lett. {\bf B114} (1982) 39-41.
  
\bibitem{Brax:2009ey}
  Brax, P., Burrage, C., Davis, A.~C., Seery, D., \& Weltman, A.,
  arXiv:0911.1267 [hep-ph]
\bibitem{Khoury:2003aq}
  Khoury, J., \& Weltman, A.,
  Phys.\ Rev.\ Lett.\  {\bf 93} (2004) 171104
  [arXiv:astro-ph/0309300]
\bibitem{Khoury:2003rn}
  Khoury, J., \& Weltman, A.,
  Phys.\ Rev.\  D {\bf 69} (2004) 044026
  [arXiv:astro-ph/0309411]  
\bibitem{Burrage:2008ii}
  Burrage, C., Davis, A.~C., \& Shaw, D.~J.,
  Phys.\ Rev.\  D {\bf 79} (2009) 044028
  [arXiv:0809.1763 [astro-ph]]
\bibitem{Brax:2004qh}
  Brax, P., van de Bruck, C., Davis, A.~C., Khoury, J., \& Weltman, A.,
  Phys.\ Rev.\  D {\bf 70} (2004) 123518
  [arXiv:astro-ph/0408415]
\bibitem{Brax:2007ak}
  Brax, P., van de Bruck, C., \& Davis, A.~C.,
  Phys.\ Rev.\ Lett.\  {\bf 99} (2007) 121103
  [arXiv:hep-ph/0703243]
\bibitem{Raffelt:1987im}
  Raffelt, G., \& Stodolsky, L., 
  Phys.\ Rev.\  D {\bf 37} (1988) 1237.

\bibitem{Barrow:1997mj}
  Barrow, J.~D., Ferreira, P.~G., \& Silk, J.,
  Phys.\ Rev.\ Lett.\  {\bf 78} (1997) 3610
  [arXiv:astro-ph/9701063]
\bibitem{Blasi:1999hu}
  Blasi, P., Burles, S., \& Olinto, A.~V.,
  Astrophys.\ J.\  {\bf 514} (1999) L79
  [arXiv:astro-ph/9812487]
\bibitem{Kronberg:1993vk}
  Kronberg, P.~P.,
  Rept.\ Prog.\ Phys.\  {\bf 57} (1994) 325.
  
\bibitem{Davis:2009vk}
  Davis, A.~C., Schelpe, C.~A.~O., \& Shaw, D.~J.,
  Phys.\ Rev.\  D {\bf 80} (2009) 064016
  [arXiv:0907.2672 [astro-ph.CO]]
\bibitem{Schelpe:2010he}
  Schelpe, C.~A.~O., arXiv:1003.0232 [astro-ph.CO]
 
 
 \bibitem{Peebles:1994xt}
  Peebles, P.~J.~E., {\it  Princeton, USA: Univ. Pr. (1993) 718 p}
\bibitem{Csaki1} Cs{\'a}ki, C.,
Kaloper, N., \& Terning, J. 2002, Physics Letters B, 535, 33
\bibitem{Grossman:2002by}
  Grossman, Y., Roy, S., \& Zupan, J.,
  Phys.\ Lett.\  B {\bf 543} (2002) 23
  [arXiv:hep-ph/0204216]

\bibitem{Mirizzi:2009aj}
  Mirizzi, A., \& Montanino, D.,
  JCAP {\bf 0912} (2009) 004
  [arXiv:0911.0015 [astro-ph.HE]]
  \bibitem{Burrage:2009mj}
  Burrage, C., Davis, A.~C., and Shaw, D.~J.,
  Phys.\ Rev.\ Lett.\  {\bf 102} (2009) 201101
  [arXiv:0902.2320 [astro-ph.CO]] 
    
\bibitem{MirRafSerp}
  Mirizzi, A., Raffelt, G.~G. \& Serpico, P.~D. 2005,
   Phys.\ Rev.\  D {\bf 72} 023501
  [arXiv:astro-ph/0506078]

\bibitem{Ostman:2004eh}
  Ostman, L., \& Mortsell, E.,
  JCAP {\bf 0502} (2005) 005
  [arXiv:astro-ph/0410501] 
  
\bibitem{Mortsell1} M{\"o}rtsell, E., \& Goobar, A.  2003, Journal of Cosmology and Astro-Particle Physics, 9, 9
\bibitem{Mortsell2} M{\"o}rtsell, E., \& Goobar, A.  2003, Journal of Cosmology and Astro-Particle Physics, 4, 3
 
\bibitem{Holdom:1985ag} 
 Holdom, B., 
{\em Phys. Lett.} {\bf B166} (1986) 196.
\bibitem{Batell:2005wa}
Batell, B., \& Gherghetta, T.,  
{\em Phys. Rev.} {\bf D73} (2006) 045016 [hep-ph/0512356]
\bibitem{Bruemmer:2009ky}
Brummer, F., Jaeckel, J., \& Khoze, V.~V., 
{\em JHEP} {\bf 06} (2009) 037 [arXiv:0905.0633]
\bibitem{Abel:2006qt}
Abel, S.~A., Jaeckel, J., Khoze, V.~V., \& Ringwald, A., 
  {\em Phys. Lett.} {\bf B666} (2008) 66--70 [hep-ph/0608248]
\bibitem{Abel:2008ai}
Abel, S.~A., Goodsell, M.~D.,  Jaeckel, J., Khoze, V.~V., \& Ringwald, A., 
  {\em JHEP} {\bf 07} (2008) 124 [arXiv:0803.1449]
\bibitem{Dienes:1996zr}
Dienes, K.~R., Kolda, C.~F., \& March-Russell, J., 
{\em Nucl. Phys.} {\bf B492} (1997) 104--118 [hep-ph/9610479]
\bibitem{Abel:2003ue}
Abel, S.~A., \& Schofield, B.~W. 
 {\em Nucl. Phys.} {\bf B685} (2004) 150--170 [hep-th/0311051]
\bibitem{Mark}
Goodsell, M.~D., Jaeckel, J., Redondo, J., \& Ringwald, A., 
 {\em DESY 09-123; DCPT/09/124; IPPP/09/62} [arXiv:0909.0515]
\bibitem{Melchiorri:2007sq}
Melchiorri, A., Polosa, A., \& Strumia, A., 
{\em Phys. Lett.} {\bf B650} (2007) 416--420
[hep-ph/0703144]
\bibitem{Gies:2006ca}
Gies, H., Jaeckel, J., \& Ringwald, A.,
  {\em Phys. Rev. Lett.}
  {\bf 97} (2006) 140402 
  [hep-ph/0607118]
\bibitem{Ahlers:2007rd}
Ahlers, M., Gies, H., Jaeckel, J., Redondo, J., \& Ringwald, A.,
 {\em Phys. Rev.} {\bf D76} (2007) 115005 [arXiv:0706.2836]
\bibitem{Burrage:2009yz}
  Burrage, C., Jaeckel, J., Redondo, J., \& Ringwald, A., 
  JCAP {\bf 0911}, 002 (2009)
  [arXiv:0909.0649 [astro-ph.CO]]

\bibitem{Ahlers:2007qf}
Ahlers, M., Gies, H., Jaeckel, J., Redondo, J., \& Ringwald, A., 
{\em Phys. Rev.} {\bf D77} (2008)  095001 [arXiv:0711.4991]
\bibitem{Jaeckel:2009dh}
Jaeckel, J., 
  {\em Phys. Rev. Lett.} {\bf 103} (2009) 080402 [arXiv:0904.1547]
\bibitem{Gies:2006hv}
Gies, H., Jaeckel, J., \&, Ringwald, A., 
{\em Europhys. Lett.} {\bf 76} (2006) 794--800 [hep-ph/0608238]

\bibitem{BOSS}
  Schlegel, D., White, M., \& Eisenstein, D.,  [with input from the SDSS-III
                  collaboration and with input from the SDSS-III],
  [arXiv:0902.4680 [astro-ph.CO]]
\bibitem{SPACE}
  Cimatti, A., {\it et al.},
  Exper.\ Astron.\  {\bf 23} (2009) 39
  [arXiv:0804.4433 [astro-ph]]
\bibitem{DUNE}
  Refregier, A., Douspis, M., \&  t.~D.~collaboration,
  arXiv:0807.4036 [astro-ph]
\bibitem{SeoEisen}
  Seo, H.~J., \& Eisenstein, D.~J.,
  Astrophys.\ J.\  {\bf 665} (2007) 14
  [arXiv:astro-ph/0701079]

\bibitem{DETF} Albrecht, A., {\it et al.} 2006, arXiv:astro-ph/0609591 


  
\end{thebibliography}
\end{document}